\documentclass[10pt]{iopart} 
\usepackage{graphicx}% Include figure files 
\usepackage{bm}      % bold math   
\usepackage{epsfig} 
\usepackage{wasysym}
\usepackage{cases}
\usepackage{fullpage}
\usepackage{setspace}

\newcommand{\simlt}%
        {\,\hbox{\lower0.6ex\hbox{$\sim$}\llap{\raise0.6ex\hbox{$<$}}}\,}

\begin{document}

\title[Velocity and processivity of helicases]{Velocity and
  processivity of helicase unwinding of double-stranded nucleic acids}

\author{M. D.  Betterton\dag, and F. J\"ulicher\ddag} \address{\dag
  Department of Applied Mathematics, University of Colorado at
  Boulder, 526 UCB, Boulder, Colorado 80309, USA} \address{\ddag
  Max-Planck Institute for the Physics of Complex Systems,
  N\"othnitzerstrasse 38, 01187 Dresden, Germany}

\begin{abstract}
  Helicases are molecular motors which unwind double-stranded nucleic
  acids (dsNA) in cells. Many helicases move with directional bias on
  single-stranded (ss) nucleic acids, and couple their directional
  translocation to strand separation.  A model of the coupling between
  translocation and unwinding uses an interaction potential to
  represent passive and active helicase mechanisms. A passive helicase
  must wait for thermal fluctuations to open dsNA base pairs before it
  can advance and inhibit NA closing.  An active helicase directly
  destabilizes dsNA base pairs, accelerating the opening rate.  Here
  we extend this model to include helicase unbinding from the
  nucleic-acid strand.  The helicase processivity depends on the form
  of the interaction potential.  A passive helicase has a mean
  attachment time which does not change between ss translocation and
  ds unwinding, while an active helicase in general shows a decrease
  in attachment time during unwinding relative to ss translocation.
  In addition, we describe how helicase unwinding velocity and
  processivity vary if the base-pair binding free energy is changed.
\end{abstract}

\section{Introduction}

Helicases are motor proteins which separate the two strands of helical
double-stranded nucleic acids (NA). Both DNA, RNA, and DNA-RNA hybrid
helicases are found in cells. Strand separation requires breaking the
base-pairing interactions between the two strands and therefore
requires energy input.  Unwinding is fueled by NTP hydrolysis,
typically of ATP.  Helicases play a role in nearly every cellular
process which involves NA, including DNA replication and repair,
recombination, transcription, translation, and RNA
processing\cite{loh96}. Aberrant functioning of helicases is
associated with genome instability (the accumulation of damage and
errors in the genome), premature aging, and cancer\cite{bra00}.

The essential common feature of all helicases is their ability to move
along NA strands and couple motion to strand separation. (We use the
terms opening, strand separation, and unwinding interchangeably.)  For
this reason, helicases are also NA translocases and share some
features with other proteins which move on NA
strands\cite{von01,sin02}.

The unwinding velocity and processivity of helicases are important for
helicase function. The unwinding velocities of helicases range from
tens to thousands of base pairs per second\cite{loh96}.  Helicase
processivity is also variable.  Processivity is most often defined as
the average number of base pairs unwound per helicase binding event.
Measured values of helicase processivity range from tens to tens of
thousands of base pairs\cite{loh96}. The processivity can be
significantly altered by accessory
proteins\cite{boehm98,soult99,noirot02}, polymerases\cite{stano05} and
multiple copies of the same helicase\cite{levin04,byrd04,tacket05}.
Attachment of an enzyme either to a surface or another protein tends
to increase the processivity\cite{yin99}.  In single-molecule
experiments on RecBCD helicase, a lower processivity was measured for
a helicase that was free in solution\cite{bianc01} than for a helicase
attached to the surface\cite{per04}.
The velocity and processivity of helicase proteins are thought to be
related to their biological roles. For example, replicative helicases
are responsible for unwinding all cellular DNA during DNA replication.
It has been proposed that replicative helicases should therefore have
a high velocity and processivity, so that a small number of helicases
can efficiently function in DNA copying. By contrast, a helicase which
functions in DNA damage repair may only need to unwind a small region
of DNA near a damage site; therefore its velocity and processivity may
be low.  Understanding the physical basis of the velocity and
processivity will help illuminate how these proteins may be optimized
for different cellular roles.

Different definitions of processivity have been given in the
literature.  They include (\textit{i}) the average time a motor stays
attached to its track, (\textit{ii}) the average number of steps a
motor travels before detaching, (\textit{iii}) the number of ATP
molecules hydrolyzed before the motor falls off the track, and
(\textit{iv}) the probability that a motor takes one more step (as
opposed to unbinding before the next step).  For helicases and other
motors which interact with obstacles, this picture is complicated.  We
may consider both the \textit{translocation processivity}, which is
the average number of forward steps taken by the helicase taken during
one binding event, and the \textit{unwinding processivity}, which is
the average number of NA base pairs (bp) unwound during a helicase
binding event. The translocation and unwinding processivity are not
equivalent, and their values vary with initial conditions. Unwinding
processivity is usually measured for helicases; however, the measured
value can depend on where the motor binds relative to where it begins
unwinding the NA.

Because helicase proteins are involved in a wide range of cellular
processes, many different types of helicase protein are found in
cells. Even the relatively simple bacterium \textit{E. coli} has at
least 11 different types of helicase. Helicases are structurally
diverse, and not all helicases share a common mechanism. However, many
helicases share the ability to translocate directionally on
single-stranded (ss) NA, especially members of superfamilies I and
II\cite{sin02}. Motion on ssNA is analogous to the motion of a
classical motor protein which moves with directional bias on an
infinite, one-dimensional lattice (figure 1). Single-stranded NA is
polar, with one end labeled the 3$'$ end and the other the $5'$ end.
Helicases which directionally translocate on ssNA are referred to as
$3'\to 5'$ helicases or $5'\to 3'$ helicases, depending on which
direction they move on the single strand. When the helicase is near
the ss-double strand (ds) junction, the helicase can move the junction
forward, creating additional ssNA ``track'' as it moves.

A natural question in the study of helicases is how a protein may
efficiently couple translocation to unwinding.  In the biochemical
literature on helicases, this coupling is classified as passive or
active\cite{lohman93,loh96,von01,sin02}. A passive helicase waits for a
thermal fluctuation that opens part of the dsNA, and then moves
forward, binds to the newly available ssNA, and prevents the NA from
closing. An active helicase directly destabilizes the dsNA, presumably
by changing the free energy of the ds state.

\begin{figure}[t]\centering
  \includegraphics[height=4cm]{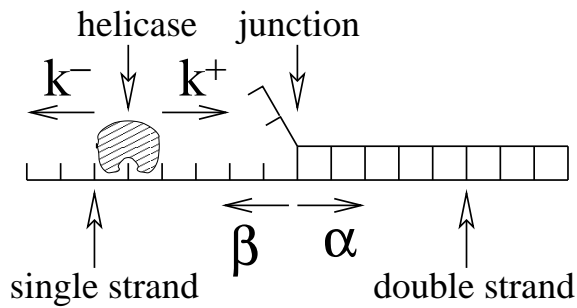}
  \caption[Sketch of helicase on a nucleic acid strand. 
  The helicase moves forward---toward the junction where the dsNA
  strand opens---at rate $k^+$ and backward at rate $k^-$. The NA
  opens at rate $\alpha$ and closes at rate $\beta$.]  {Sketch of
    helicase on a nucleic acid strand.  The helicase moves
    forward---toward the junction where the dsNA strand opens---at
    rate $k^+$ and backward at rate $k^-$. The NA opens at rate
    $\alpha$ and closes at rate $\beta$.}
\label{hopdna} 
\end{figure}

Many different types of helicase have been studied experimentally in
some detail\cite{loh96,von01,del02}.  In recent years, single-molecule
experiments have been performed for several helicases, directly
measuring helicase unwinding rates. The helicase RecBCD, which is also
an exonuclease (it degrades one of the unwound strands as it moves)
has been studied by several
groups\cite{dohon01,bianc01,spies03,per04,handa05}.  Interpretation of
the RecBCD data is complicated by the fact that RecBCD contains two
helicases of opposite polarity, which move on both strands of the
DNA\cite{taylor03,dillin03,singl04}.  Bulk kinetic experiments on
RecBCD have tried to determine the kinetic step size by fitting the
data to a model of helicase
unwinding\cite{lucius04,lucius04b,fisch04,lucius02}; to date physical
steps of this helicase have not been directly observed.  Single-strand
translocation and double-strand unwinding were observed for Rep
helicase using single-molecule FRET\cite{ha02,rasnik04}.  The motion
of the RNA helicase DbpA has been observed with AFM\cite{hen01}.
Single-molecule experiments on UvrD helicase have observed unwinding,
and evidence of strand-switching\cite{dessin04}.  Finally,
single-molecule FRET has been used to study the T4 replisome, the
replication machinery which includes the T4 helicase\cite{xi05}.

Although extensive biochemical and structural studies of helicases
have been performed, few descriptions of the physics of helicase
unwinding exist. Previous work includes a ``flashing field'' model
specific to hexameric ring helicases\cite{doe95}, a description of a
helicase as a biased random walk, which considered how the density of
histones affects the random walk\cite{che97}, and a model where
helicase motion is represented as a propagating
front\cite{bha03,bha04}. Molecular dynamics simulations have addressed
properties of PcrA helicase\cite{cox03}.  Recent work includes a
proposal that HCV helicase functions by a ratchet
mechanism\cite{levin05}, a theory for two coupled motor proteins,
proposed to describe RecBCD helicase\cite{stukal05}, and a detailed
description of the mechanochemistry of T7 helicase\cite{liao05}.  A
physical description of helicase unwinding of NA has been proposed
which contains both active and passive opening as different cases in a
general framework\cite{bet03,bet05}.  Here we extend the description
of helicase unwinding to allow calculation of helicase processivity.

We represent a helicase by a particle which moves with directional
bias on a one-dimensional lattice (corresponding to the ssNA). The
ss-ds junction represents an obstacle on the lattice which blocks the
helicase motion while also moving due to thermal fluctuations. Our
model describes different forms of the interaction between the
particle and the mobile obstacle.  This scenario---a moving particle
which interacts with an obstacle---also exists in other biological
systems. Peskin, Odell, and Oster introduced the ``polymerization
ratchet'' to describe how a growing biological polymer can exert a
force against a fluctuating obstacle\cite{pes93}.  They argued that
the rate of polymerization is limited by the time required for the
obstacle to diffuse one monomer size.  In the language of our model,
described below, this scenario corresponds to a hard-wall
interaction potential between the growing tip and the obstacle. As we
show here, other forms of interaction can show significantly different
velocity and processivity.  Another example of the interaction of a
motor with a second degree of freedom is the kinesin-family motor
protein MCAK, which induces microtubule depolymerization if it
interacts with the microtubule end\cite{hunter03,bring04,klein05}.
This process is relevant in the generation of force during the
separation of chromosomes by the mitotic spindle.  Finally, the
interaction of two motor proteins on a filament is another example for
which our approach is relevant.

In this paper, we first review a simple description which captures
both active and passive unwinding by helicases\cite{bet03,bet05}. This
framework considers a single helicase which does not unbind from the
NA strand, and permits us to calculate the unwinding velocity for a
given interaction potential (section \ref{oldmodel}). The unwinding
velocity for a passive helicase with hard-wall interaction potential
is typically significantly slower than the ss translocation rate of
the motor far from the ss-ds junction. For a simple form of active
opening, the velocity approaches the single-strand translocation rate
of the motor. In other words, an optimized active helicase can unwind
NA as fast as it translocates on ssNA.  Comparable rates of ss
translocation and unwinding is therefore a signature of active
opening.

We then extend this description to include a nonzero unbinding rate,
and develop a simple model of unbinding that captures the key effects.
We assume that the unbinding rate may be different for a helicase
translocating on ssNA than for a helicase unwinding dsNA, because the
interaction potential alters the unbinding rate when the helicase is
near the ss-ds junction. If the interaction free energy is larger,
than the helicase will unbind more quickly. In particular, we assume
that the free energy difference between the bound state and the
barrier to unbinding is decreased by the amount of the interaction
potential. 
We assume that the unbinding rate is independent of the helicase
biochemical state.  The unbinding rate is determined only by the
height of the energy barrier separating the bound from the unbound
states.

This model of unbinding allows us to calculate different measures of
processivity (section \ref{newmodel}).  We discuss the dependence of
the processivity on the shape of the interaction potential.  The
average attachment time of a helicase is particularly simple for a
passive helicase: the helicase unbinding rate during unwinding is
equal to the unbinding rate during ss translocation (section
\ref{passive}). By contrast, the helicase attachment time is lower
during unwinding than during ss translocation for an active helicase
(section \ref{active}).  Decreased attachment time during unwinding is
a signature of active unwinding.

Finally, in our conclusion, we discuss the connection
between our results and experiments (section \ref{conclusion}).

\section{Active and passive unwinding}
\label{oldmodel}

In our simplified discrete description of helicase unwinding, the
position of the motor---the helicase---along its track is labeled by
the integer $n$, and the position of the obstacle---the ss-ds
junction---is denoted $m$ (figure \ref{hopdna}).  The motor is assumed
to move toward increasing $n$, and we expect $n\le m$.  The
interaction between motor and obstacle is characterized by an
interaction energy $U(m-n)$ which depends only on the obstacle-motor
separation.  We assume $U \to 0$ for $m \gg n$ and $U \to \infty$ for
$m < n$.

One simple form of this potential is a hard-wall interaction of zero
range: $U=0$ for $m>n$ and $U=\infty$ for $m\leq n$.  For a hard-wall
potential, a motor which is near the obstacle ($n=m-1$) can advance
only if the obstacle undergoes a fluctuation which increases $m$. This
situation corresponds to passive unwinding.  For active unwinding, the
interaction between the motor and obstacle has nonzero range: when the
helicase is near the ss-ds junction, both the kinetics of NA opening
and the helicase motion are altered by the interaction.
We will discuss interaction potentials which represent active
unwinding below.

We denote the rates of forward and backward hopping of the motor on
the lattice far from the obstacle by $k^+$ and $k^-$.  The hopping
rates for forward and backward motion of the junction (corresponding
to the opening and closing of dsNA) in the absence of the helicase are
denoted $\alpha$ and $\beta$.  The ratio of these rates is given by
$\alpha/\beta=e^{-\Delta G}$ where $\Delta G$ is the free energy
difference per base between dsNA and two complementary ssNA strands.
(Note that we use units where $k_B T=1$.)  This energy difference is
positive when dsNA is thermodynamically stable.  For simplicity, we
write a similar expression for the ratio of forward and backward
hopping rates of the helicase, $k^+/k^-=e^{\Delta\mu}$.  Here
$\Delta\mu$ denotes the chemical free energy of ATP hydrolysis which
drives helicase motion. This expression applies when each hydrolysis
event is tightly coupled to a forward step on the NA. We use this
simplification here to discuss principles which do not depend on the
validity of this assumption.  Note that an effective value of $\Delta
\mu$ can be derived in certain limits from a more detailed description
of the mechanochemical coupling\cite{bet05}.

The interaction between the helicase and the ss-ds junction modifies
the hopping rates.  We express the ratios of these rates as
\begin{subequations}
\begin{eqnarray} 
\label{posdependence} 
\frac{\beta_{j}}{\alpha_{j-1}} &= \frac{\beta}{\alpha} 
        e^{-[U(j-1)-U(j)]}, \\
\frac{k^+_{j}}{k^-_{j-1}} &= \frac{k^+}{k^-} 
        e^{-[U(j-1)-U(j)]} .
\label{posdepk} 
\end{eqnarray} 
\end{subequations}
where $\alpha_j$, $\beta_j$, $k^+_j$ and $k^-_j$ are the
position-dependent rates when the helicase and the junction are
separated by $j=m-n$ bases.  To fully specify the position-dependent
rates we write
\begin{subequations}
\label{ratedef}
\begin{eqnarray}  
k^+_{j}= k ^ +e^{-f[U(j-1)-U(j)]}, \\  
k^-_{j-1}=k ^-e^{-(f-1)[U(j-1)-U(j)]}, \\  
\beta_{j}=\beta e^{- f[U(j-1)-U(j)]},\\  
\alpha_{j-1} = \alpha e^{-(f-1)[U(j-1)-U(j)] } .  
\end{eqnarray} 
\end{subequations}
where the parameter $f$ describes the energy barrier of the
transitions.  For a one-dimensional reaction, $f$ corresponds to the
fractional distance of the peak of the barrier between the two
adjacent states and thus $0<f<1$.

\begin{figure}[t] 
  \centering \includegraphics[height=3cm]{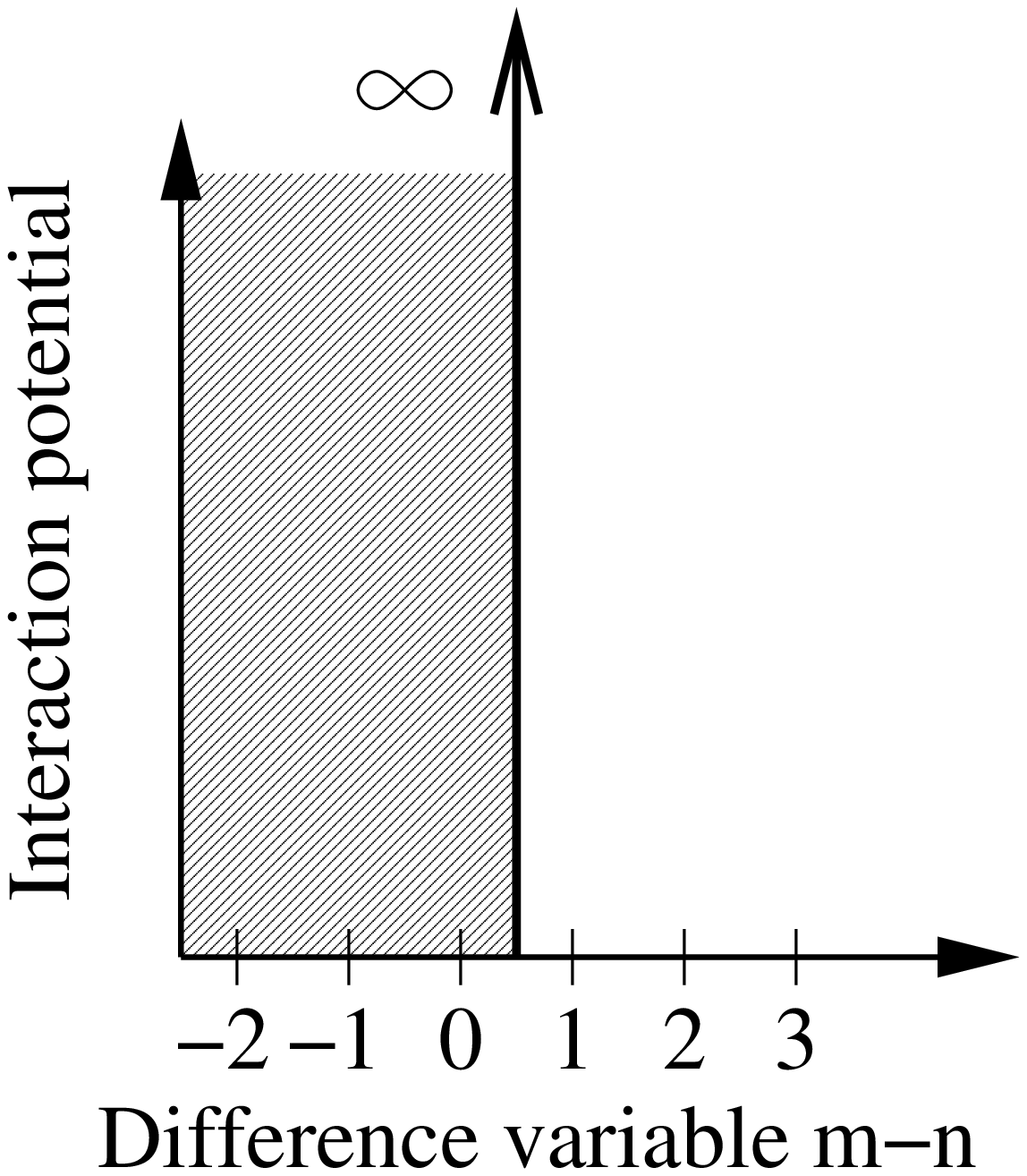}
  \includegraphics[height=3cm]{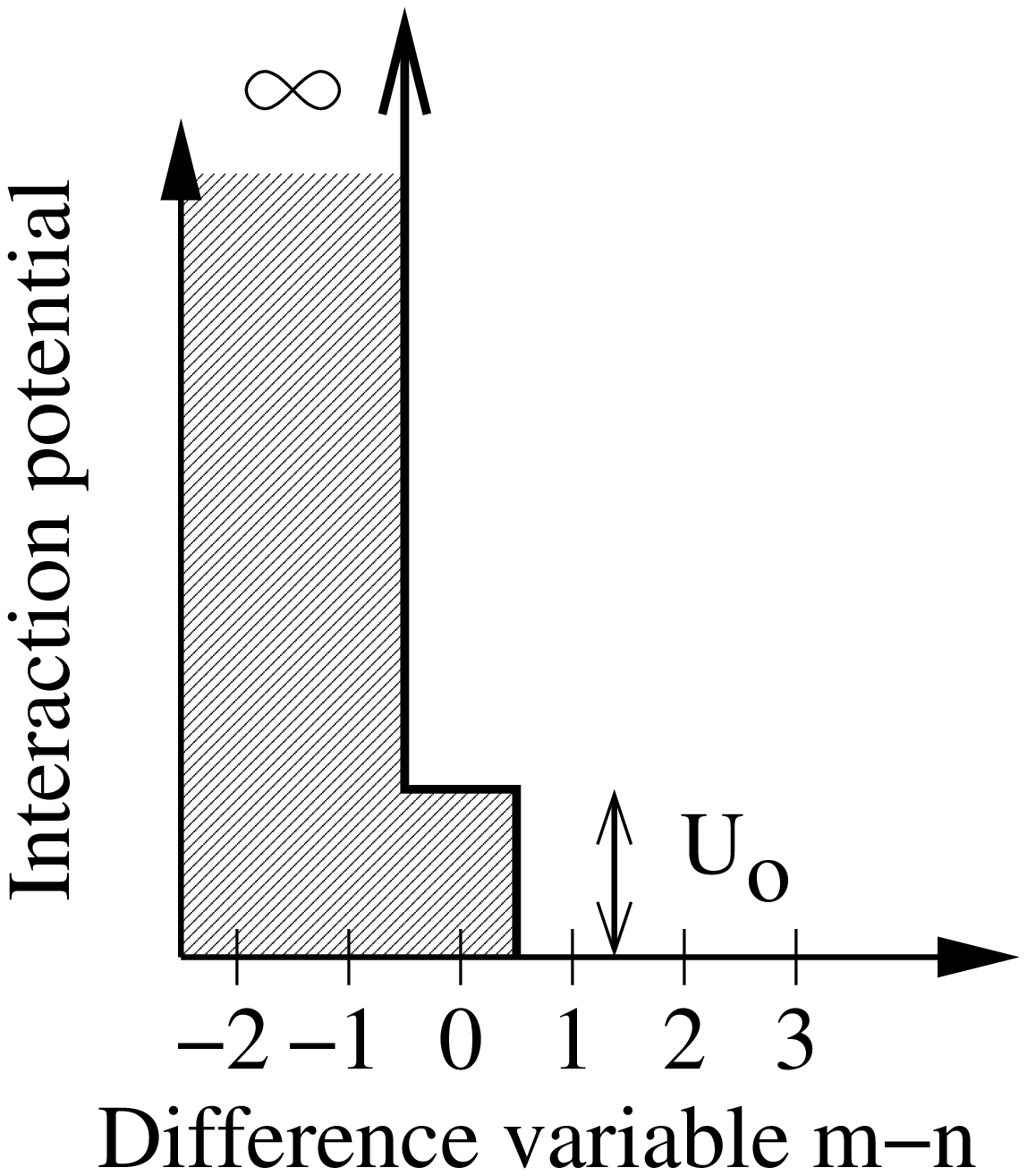}
  \includegraphics[height=3cm]{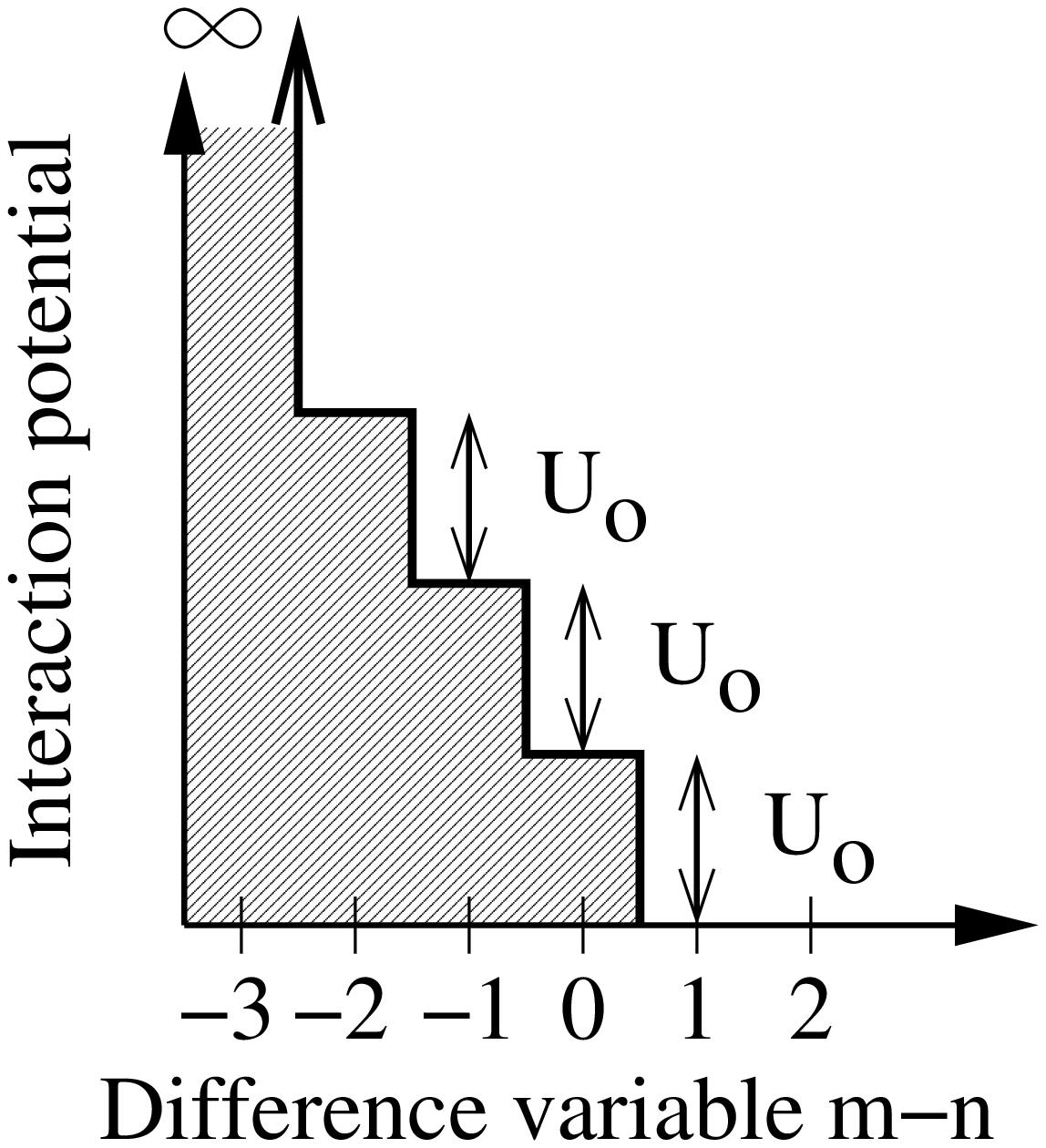}
\caption {Interaction energies between the helicase and ss-ds
  junction as a function of their distance $m-n$ in number of bases.
  (a) Hard-wall potential. (b) Potential with a single step. (c)
  Potential with three steps.}
\label{enplot} 
\end{figure}
 
For the simple case of a hard-wall potential (figure \ref{enplot}a),
the helicase prevents NA closing when $j=1$. Therefore $\beta_{1}=0$
because $U(0)$ is infinite. Infinite $U(0)$ also implies $k^+_{1}=0$:
the helicase must wait until a thermal fluctuation opens the NA before
advancing.  An interaction potential with nonzero range corresponds to
enzymatically assisted opening. For simplicity, we use a linear
potential with range $N$ and slope $U_o$.  The discretized potential
is shown in figures \ref{enplot}b and \ref{enplot}c. The potential
energy increases in $N$ steps, each of energy $U_0$, before a hard
wall is reached.  The increase in energy due to this potential
facilitates opening of the dsNA: $U(j-1)>U(j)$ implies
$\beta_{j}/\alpha_{j-1}<\beta/\alpha$.  In addition, the interaction
energy slows helicase forward motion.  We assume $U \to 0$ for $j \to
\infty$, so that no interaction occurs for large separations.

The probability $P(j,l,t)$ that the helicase and junction are at
separation $j$ and midpoint position $l=m+n$ satisfies the master
equation
\begin{eqnarray}
\frac{d P(j,l,t)}{d t} = &-(\alpha_{j} +\beta_{j}  
        +k^+_{j} +k^-_{j}) P(j,l) + \alpha_{j-1} P(j-1,l-1)  \nonumber
        \\
        &+  
        \beta_{j+1} P(j+1,l+1) 
        +k^+_{j+1} P(j+1,l-1) +k^-_{j-1} P(j-1,l+1). \label{basicprob}  
\end{eqnarray}  
Since the rates depend only on $j$, we define the distribution ${\cal
  P}_j=\sum_l P(j,l,t)$.  After a relaxation time which depends on the
rates $\alpha$, $\beta$, $k^+$, and $k^-$, the distribution ${\cal
  P}_j$ relaxes to a stationary state which satisfies the recursion
relation
\begin{equation}  
{\cal P}_{j+1} =  \frac{k^-_{j}+\alpha_{j}}{k^+_{j+1}+\beta_{j+1}}
{\cal P}_{j}. 
\label{nocurr}  
\end{equation}  
This steady state distribution can be used to calculate the mean
velocity (bp s$^{-1}$)\cite{bet03,bet05}
\begin{equation}  
v = \frac{1}{2} \sum_j (k^+_{j} +\alpha_{j}-k^-_{j} -\beta_{j}){\cal P}_j.  
\label{vel}  
\end{equation}
This expression for $v$ has a simple physical interpretation---the
quantity in parentheses is the unwinding rate at separation $j$, which
is multiplied by the probability ${\cal P}_j$ of finding the complex
at separation $j$. The effective diffusion coefficient which
characterizes velocity fluctuations is
\begin{equation}  
D=\frac{1}{4} \sum_j (k^+_{j} +\alpha_{j}+k^-_{j} +\beta_{j}){\cal P}_j.  
\end{equation}  
We assume that the helicase remains bound to the NA for all time.
Therefore the expressions for $v$ and $D$ are true steady-state values
reached by a real system in the long-time limit.  In the results
below, unwinding velocities of a bound helicase are calculated under
this steady-state assumption.

\section{Processive and unprocessive unwinding}
\label{newmodel}

In section \ref{oldmodel} we assumed that the motor never unbinds from
the track, so the model represents a motor which is infinitely
processive. Real motors have finite processivity---they unbind
eventually. This effect can be incorporated in our description by
introducing the rate of unbinding $\gamma$. We assume that $\gamma$
depends on the separation $j$ between motor and obstacle, but has no
other position or time dependence.  For a repulsive potential such as
the linear potential discussed above, when the motor and obstacle are
close to each other the free energy of motor and junction is
increased. This typically leads to an increased rate of unbinding.
This effect can be described by writing\cite{parmeg01}
\begin{equation} 
\gamma_j=\gamma e^{U(j)}.
\end{equation} 
Here $\gamma>0$ is the detachment rate for motion on a single 
strand far from the junction.

The time evolution of the probability distribution is then described by
\begin{eqnarray}  
\fl \frac{d P(j,l)}{d t} = -( k^+_j +k^-_j 
        +\alpha_j +\beta_j +\gamma_j) P(j,l) + \alpha_{j-1} P(j-1,l-1) +  
        \beta_{j+1} P(j+1,l+1) \nonumber\\  
+k^+_{j+1} P(j+1,l-1) +k^-_{j-1} P(j-1,l+1).
\label{prob}  
\end{eqnarray} 
Note that this expression differs from equation (\ref{basicprob}) only
in the additional term proportional to the unbinding rate $\gamma_j$.
The other rates are unchanged as defined above, in equation
(\ref{ratedef}).  We assume that at time $t=0$, the helicase and
junction have a specific position determined by $j=j_o$ and $l=l_o$.
Therefore,
\begin{equation} P(j,l,t=0)=\delta_{j j_o}\delta_{l l_o},
\label{initc}
\end{equation}
where $\delta_{jk}=1$ for $j=k$ and $0$ otherwise.

\subsection{Processivity and attachment time} 

We characterize the processivity by the average attachment time
$\langle \tau \rangle$ of the motor and by both the
\textit{translocation processivity} $\langle \delta n \rangle$, which
is the average number of forward steps taken by the helicase during
one binding event, and the \textit{unwinding processivity} $\langle
\delta m \rangle$, which is the average number of NA bp unwound during
one helicase binding event.  These measures of processivity depend on
initial conditions. If the helicase binds near the ss-ds junction,
then the translocation and unwinding processivities are similar.
However, if the helicase binds to a ss region far from the junction,
then the unwinding processivity can be much smaller than the
translocation processivity. If the NA can close on average under these
conditions, the corresponding unwinding processivity becomes negative.

The mean attachment time is defined as
\begin{equation} 
\langle\tau\rangle = \int_0^{\infty}dt\ t \sum_{j,l} \gamma_j P(j,l,t).
\label{tau}
\end{equation} 
Similarly, the translocation and unwinding processivities can be
defined by
\begin{eqnarray} 
\label{deln}
\langle\delta n\rangle = \frac{\langle\delta l\rangle- \langle\delta
  j\rangle}{2}, \\
\langle\delta m\rangle = \frac{\langle\delta l\rangle+ \langle\delta
  j\rangle}{2}. 
\label{delm}
\end{eqnarray} 
Here, the average changes in $j$ and $l$ during unwinding are
\begin{eqnarray} 
\label{delj}  
\langle\delta j\rangle &= \int_0^{\infty}dt\ \sum_{j,l} (j-j_o) \
  \gamma_j P(j,l,t),\\
  \langle\delta l\rangle &= \int_0^{\infty}dt\ \sum_{j,l} (l-l_o) \
  \gamma_j P(j,l,t).
\label{dell2}
\end{eqnarray} 
The translocation processivity $\langle \delta n \rangle = \langle
n-n_o \rangle$ is the mean number of steps the motor takes in one
direction before detaching and the unwinding processivity $\langle
\delta m \rangle = \langle m-m_o \rangle$ is the mean number of steps
the obstacle moves per binding event.  The above expressions can be
simplified using Laplace transforms\cite{parmeg01}.  The
Laplace-transform of $P(j,l,t)$ is
\begin{equation} 
\tilde{P}_{j,l}(s)=\int_0^{\infty} dt \ e^{-s t}\ P(j,l,t).
\label{laplacet}
\end{equation} 
%We rewrite equation (\ref{prob}) as
%\begin{equation} 
%\frac{d P(j,l,t)}{d t} = -\gamma_j P(j,l,t) -\Delta I_j -
%        \Delta I_l, 
%\label{prob2}  
%\end{equation} 
%where $\Delta I_j=I_j-I_{j-1}$ with $I_j=\alpha_j P(j,l)-\beta_{j+1}
%P(j+1,l+1)$ and  $I_l=k^+_j P(j,l)-k^-_{j-1} P(j-1,l+1)$. 
Summing equation (\ref{prob})
%equation (\ref{prob2}) 
over $j$ and $l$, we obtain
\begin{equation} 
\sum_{j,l}\frac{d P(j,l,t)}{d t} %=\sum_{j,l}\left[ -\gamma_j P(j,l,t)
%  -\Delta I_j -       \Delta I_l \right]  
=\sum_{j,l} -\gamma_j 
P(j,l,t).
\label{prob3}
\end{equation}
%Note that the sums of $\Delta I_j$ and $\Delta I_l$ are zero when
%summed over all $j,l$.  
We can thus simplify the expression for
$\langle\tau\rangle$ in equation (\ref{tau}) by integrating by parts:
\begin{eqnarray}
  \langle \tau \rangle  &= - \sum_{j,l} \int_0^{\infty}dt\ t \frac{d
    P(j,l,t)}{d t}\nonumber \\ 
  &=  \sum_{j,l} \int_0^{\infty}dt\ P(j,l,t).
\label{tau3}
\end{eqnarray}
The boundary terms vanish because at the lower boundary $t=0$, while
for long times the probability approaches zero at all sites because
the motor unbinds.

%Now, comparing to the definition of the Laplace-transformed variable
%in equation (\ref{laplacet}), we define
Defining the coefficients
\begin{equation} 
Q_{j,l}=\tilde{P}_{j,l}(s=0)=\int_0^{\infty} dt\ P(j,l,t),
\end{equation} 
the average attachment time can be expressed as 
\begin{equation} 
\langle\tau\rangle =  \sum_{j,l}
Q_{j,l}. \label{tau4} 
\end{equation} 
Similarly, the $j$ and $l$ processivities are given by
\begin{eqnarray} 
  \langle\delta j\rangle &= \int_0^{\infty}dt\ \sum_{j,l} (j-j_o) \
  \gamma_j P(j,l,t),\\ 
 % &=  \sum_{j,l} (j-j_o)  \gamma_j \int_0^{\infty}dt\  P(j,l,t),\\
 % &=  \sum_{j,l} (j-j_o) \gamma_j \tilde{P}_{j,l}(s=0),\\
  &=  \sum_{j,l} (j-j_o) \gamma_j Q_{j,l},
\label{delj2}
\end{eqnarray} 
and 
\begin{equation} 
\langle\delta l\rangle = \sum_{j,l} (l-l_o) \gamma_j Q_{j,l}.
\label{dell}
\end{equation} 
We therefore need to calculate $Q_{j,l}$, the Laplace-transformed
probability at $s=0$, to obtain the average attachment time and the
mean change in $j$ and $l$ during a single binding event.  The
translocation and unwinding processivities are obtained by equations
(\ref{deln}) and (\ref{delm}).

The coefficients $Q_{j,l}$ are obtained by solving the
Laplace-transform of equation (\ref{prob}) for $s=0$:
\begin{eqnarray}  
  \fl  \int_0^{\infty} dt \ e^{-s t}\ \frac{d P(j,l)}{d t} = -( k^+_j
  +k^-_j +\alpha_j +\beta_j +\gamma_j) \tilde{P}_{j,l}(s) +
  \alpha_{j-1} \tilde{P}_{j-1,l-1}(s)+ \beta_{j+1}
  \tilde{P}_{j+1,l+1}(s)\nonumber\\ 
  +k^+_{j+1}\tilde{P}_{j+1,l-1}(s) +k^-_{j-1}
  \tilde{P}_{j-1,l+1}(s) .
\end{eqnarray}
Note that the rates have no explicit time dependence. As above,
integrating by parts leads to
\begin{eqnarray} 
\int_0^{\infty} dt \ e^{-s t}\ \frac{d P(j,l)}{d t} 
  &= \left. P(j,l,t)  e^{-s t} \right|^{\infty}_0+s\int_0^{\infty} dt
  \ e^{-s t} P(j,l,t),\nonumber\\
&= -P(j,l,t=0)+s\tilde{P}_{j,l}(s),\nonumber\\
&= -\delta_{jj_o}\delta_{ll_o}+s\tilde{P}_{j,l}(s).
\end{eqnarray} 
Here we have used the initial condition, equation (\ref{initc}).  Thus
the Laplace-transformed equations are
\begin{eqnarray}  
  \fl  -\delta_{jj_o}\delta_{ll_o}+s\tilde{P}_{j,l}(s) = -( k^+_j +k^-_j
  +\alpha_j +\beta_j +\gamma_j) \tilde{P}_{j,l}(s) + \alpha_{j-1}
  \tilde{P}_{j-1,l-1}(s)+ \beta_{j+1} \tilde{P}_{j+1,l+1}(s)\nonumber\\ 
  +k^+_{j+1}\tilde{P}_{j+1,l-1}(s) +k^-_{j-1} \tilde{P}_{j-1,l+1}(s).
\end{eqnarray}
In general, these equations could be solved for arbitrary $s$.  We
only require the Laplace-transformed probability evaluated at $s=0$.
The coefficients $Q_{j,l}$ satisfy the equation
\begin{equation} 
\fl \delta_{jj_o}\delta_{ll_o} =
( k^+_j +k^-_j +\alpha_j +\beta_j +\gamma_j)Q_{j,l} - \alpha_{j-1}
Q_{j-1,l-1}\\ -   
        \beta_{j+1} Q_{j+1,l+1} -k^+_{j+1}Q_{j+1,l-1}-k^-_{j-1}
        Q_{j-1,l+1}.
\label{q1}
\end{equation} 
These equations describe a discrete drift-plus-diffusion system in two
dimensions with an inhomogeneous term.

Equation (\ref{q1}) can be solved by a product ansatz
\begin{equation}
Q_{j,l}=R_j T_l
\end{equation}
Using this ansatz in equation (\ref{q1}), and choosing the
normalization $\sum_l T_l=1$ and $\sum_j \gamma_j R_j=1$, the
coefficients $R_j$ satisfy
\begin{equation}  
 -\delta_{jj_o} =-(
  k^+_j +k^-_j +\alpha_j +\beta_j +\gamma_j) R_j  +
  (\alpha_{j-1}+k^-_{j-1}) R_{j-1}  + (\beta_{j+1}+k^+_{j+1})
  R_{j+1}.
\label{q1j}
\end{equation}
The coefficients $T_l$ satisfy the equation
\begin{equation}
  \label{teq}
  \delta_{ll_o} = (p+q+1) T_l -p T_{l-1} - q T_{l+1}.
\end{equation}
Here we have defined
\begin{eqnarray}
  \label{p}
  p&=\sum_j (\alpha_j+k^+_j) R_j, \\
q&= \sum_j (\beta_j+k^-_j) R_j.
\end{eqnarray}

\subsection{Simplified expressions for processivity}
\label{1d}

The average attachment time $\langle \tau \rangle$ and the quantities
$\langle \delta j \rangle$ and $\langle \delta l \rangle$ can be
expressed in terms of the $R_j$ and $T_l$:
\begin{eqnarray} 
\label{tau2j}
\langle\tau\rangle &= \sum_{j} R_{j}  \label{tau4j}, \\
  \label{delj2j}
 \langle\delta j\rangle   &=  \sum_{j} (j-j_o) \gamma_j R_{j},\\
\label{dells}
 \langle\delta l\rangle &=  \sum_l (l-l_o) T_l. 
\end{eqnarray}

The solutions to equation (\ref{q1j}) depend on the shape of the
interaction potential. Some examples are discussed below. However, in
general equation (\ref{teq}) can be solved formally.  Equation
(\ref{teq}) is a second-order, linear, inhomogeneous difference
equation. Since we have assumed translational invariance in $l$, the
resulting solution is independent of $l_o$.  Therefore we choose
$l_o=0$ for convenience.  The solutions have the form $T_l = y^l$ for
$l \neq 0$. Using this ansatz, $y$ obeys a quadratic equation
\begin{equation}
y^2-(1+a)y+(a-b)=0, \label{qe}
\end{equation}
where $a=(1+p)/q$ and $b=1/q$.
We denote by $y_+$ and $y_-$ the positive and negative roots of equation
(\ref{qe}).  For $l\geq 0$, $T_l=A' y_-^l$, while for $l\leq 0$
the solution is $T_l=A y_+^l$, since $T_l$ must vanish for $l\to
\pm\infty$.  Requiring that $T_0$ is the same for both expressions
implies $A=A'$. Finally, equation (\ref{teq}) for $l=0$ leads to
\begin{equation}
  \label{zeroeq}
%A=\frac{b}{ \left( 1+a-y_--\frac{a-b}{y_+}\right) } .
A=b \left( 1+a-y_--\frac{a-b}{y_+}\right)^{-1}.
\end{equation}
Formally, we can use equation (\ref{dells}) to determine $\langle
\delta l \rangle$:
\begin{eqnarray}
  \langle\delta l\rangle &=  \sum_l l T_l,\\ %
%&=  A \left( \sum_{l=-1}^{\infty} l y_+^l + \sum_{l=1}^{\infty} l
%  y_-^l \right),\\
&=  A \sum_{l=1}^{\infty} \left(  l  y_-^l -l y_+^{-l} \right),\\
&=  A\left[ \frac{y_-}{(1-y_-)^2}-\frac{y_+}{(1-y_+)^2}\right].
\label{dellans}  
\end{eqnarray}
Note, however, that $A$ and $y_\pm$ depend on the shape of the
interaction potential via the coefficients $R_j$.  Once these
coefficients have been determined, the quantities $\langle \tau
\rangle$, $\langle \delta j \rangle $, and $\langle \delta l \rangle$,
and therefore $\langle \delta n \rangle$ and $\langle \delta m \rangle
$, can easily be obtained.

section{Velocity and processivity for a hard-wall interaction potential}
\label{passive}

\subsection{Velocity}

For the hard-wall potential, the unwinding velocity is given by 
\begin{equation}  
v_{HW}=\frac{\alpha  k^+ - \beta k^-}{\beta+k^+}  
\label{vhardwall}  
\end{equation}  
The velocity is positive whenever $k^+/k^- > \beta/\alpha$, that is,
the free energy change $\Delta G$ which drives NA closing must be
smaller than the chemical potential $\Delta\mu$ of ATP hydrolysis. The
maximum $v_{HW}$ occurs for a unidirectional helicase ($k^-=0$).  This
upper bound is
\begin{equation} 
v_{HW}^{\rm max}=\frac{\alpha}{\beta}\left(\frac{
k^+}{1+k^+/\beta}\right)\approx\frac{\alpha}{\beta} k^+.
\label{vhardmax} 
\end{equation}  
where the approximation holds if $k^+\ll\beta$.  Thus a passive
helicase unwinds more slowly than it translocates on ssNA by a factor
$\approx \alpha/\beta$.  This result has a simple interpretation: the
base pair adjacent to the helicase has a probability $\alpha/\beta$ of
being open. Thus, when the helicase attempts a forward hop it succeeds
with probability $\alpha/\beta$.

\subsection{Processivity}
\label{passiveg} 

For a hard-wall potential, the average attachment time, translocation
processivity, and unwinding processivity can be calculated analytically.
We solve equation
(\ref{q1j}) for $R_j$ for the simple case $j_o=1$. For $j > 1$, the rates are
independent of $j$ and we have
\begin{equation}
  R_{j+1} =(1+d)R_j+(d-e)R_{j-1},\label{r1}
\end{equation}
where we have defined
\begin{eqnarray}
  d&=\frac{\alpha+k^-+\gamma}{\beta+k^+},\\
  e&=\frac{\gamma}{\beta+k^+}.
\end{eqnarray}
This equation has solutions of the form $R_j=x_\pm^j$, where 
%\begin{equation}
%  x_{\pm}=\frac{1+d}{2}\pm \frac{1-d}{2} \sqrt{1+\frac{4
%      \epsilon}{(1-d)^2} } 
%\end{equation}
the $x_\pm$ are the
%are the 
positive and negative root of the quadratic equation
$x^2-(1+d)x+(e-d)=0$.  The boundary condition that $R_j$ vanishes for
large $j$ imposes $R_j=B x_-^j$. From equation (\ref{q1j}) at $j=1$,
we find $(\beta+k^+) (R_2-d R_1)=1$ and
\begin{equation}
  B=[x_-(\beta+k^+)(d-x_-)]^{-1}. 
\end{equation}
With these expressions, $\langle \tau \rangle = \gamma^{-1}$.
This simple result follows since for the hard wall case
$\gamma_j=\gamma$ is independent of the separation $j$.

\begin{figure}[t] 
  \centering \includegraphics[height=5.5cm]{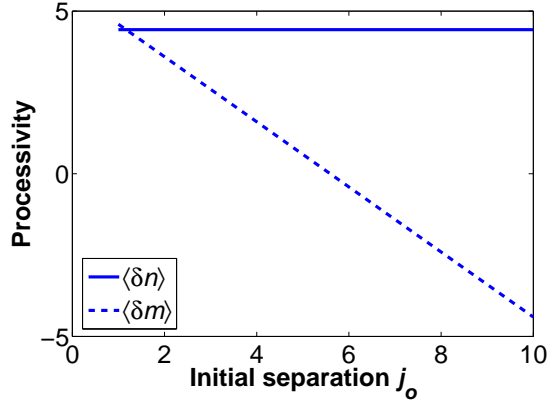}
\caption {The translocation processivity $\langle \delta n \rangle $
  and unwinding processivity $\langle \delta m \rangle $ as a function
  of the initial separation $j_o$ between motor and junction. When
  $j_o=1$, the translocation and unwinding processivities are
  approximately equal.  As the separation $j_o$ increases, the
  translocation processivity $\langle \delta n \rangle $ is
  approximately constant, while the unwinding processivity $\langle
  \delta m \rangle $ decreases.  The parameters are $\alpha=10^5$
  s$^{-1}$, $\beta=7\times 10^5$ s$^{-1}$, $k^+=1$ base s$^{-1}$,
  $k^-=0.01$ base s$^{-1}$, $\gamma=0.03$ s$^{-1}$.  Calculations were
  done numerically using a grid size of $M=100$ in $j$.}
\label{jchange} 
\end{figure} 
The average change in $j$ is
\begin{eqnarray}
  \langle \delta j \rangle %&= \sum (j-j_o) \gamma_j R_j = \gamma A
%  \sum_{j=1}^{\infty} j x_-^j - \gamma \langle \tau \rangle, \nonumber\\
&= \frac{x_-}{1-x_-}
\end{eqnarray}
To determine $\langle \delta l \rangle $, we 
note that for the hard wall case
\begin{eqnarray}
  \label{p2}
  p&=%\sum_j (\alpha_j+k^+_j) R_j=
  \gamma^{-1} (\alpha+k^+ x_-), \\
q&= %\sum_j (\beta_j+k^-_j) R_j=
\gamma^{-1} (k^-+\beta x_-).
\end{eqnarray}
The quantity $\langle \delta l \rangle $ now follows from equation
(\ref{dellans}).  In the hard-wall case, the translocation
processivity is approximately the unwinding velocity times the
attachment time: $\langle \delta m \rangle \approx v \langle \tau
\rangle$, as we would expect for position-independent detachment.
Note that the this result depends on
initial conditions and is valid for $j_o=1$.

When $j_o=1$, the translocation and unwinding processivities are
approximately equal.  As the separation between the motor and obstacle
increases, the unwinding processivity drops. A negative unwinding
processivity reflects net backwards motion of the obstacle while the
motor is bound. For a large separation between the motor and obstacle,
they have no effect on each other.  In figure \ref{jchange}, the
processivities are shown as a function of $j_o$ with other parameters
fixed.

\begin{figure}[t] 
\centering
\includegraphics[height=5.0cm]{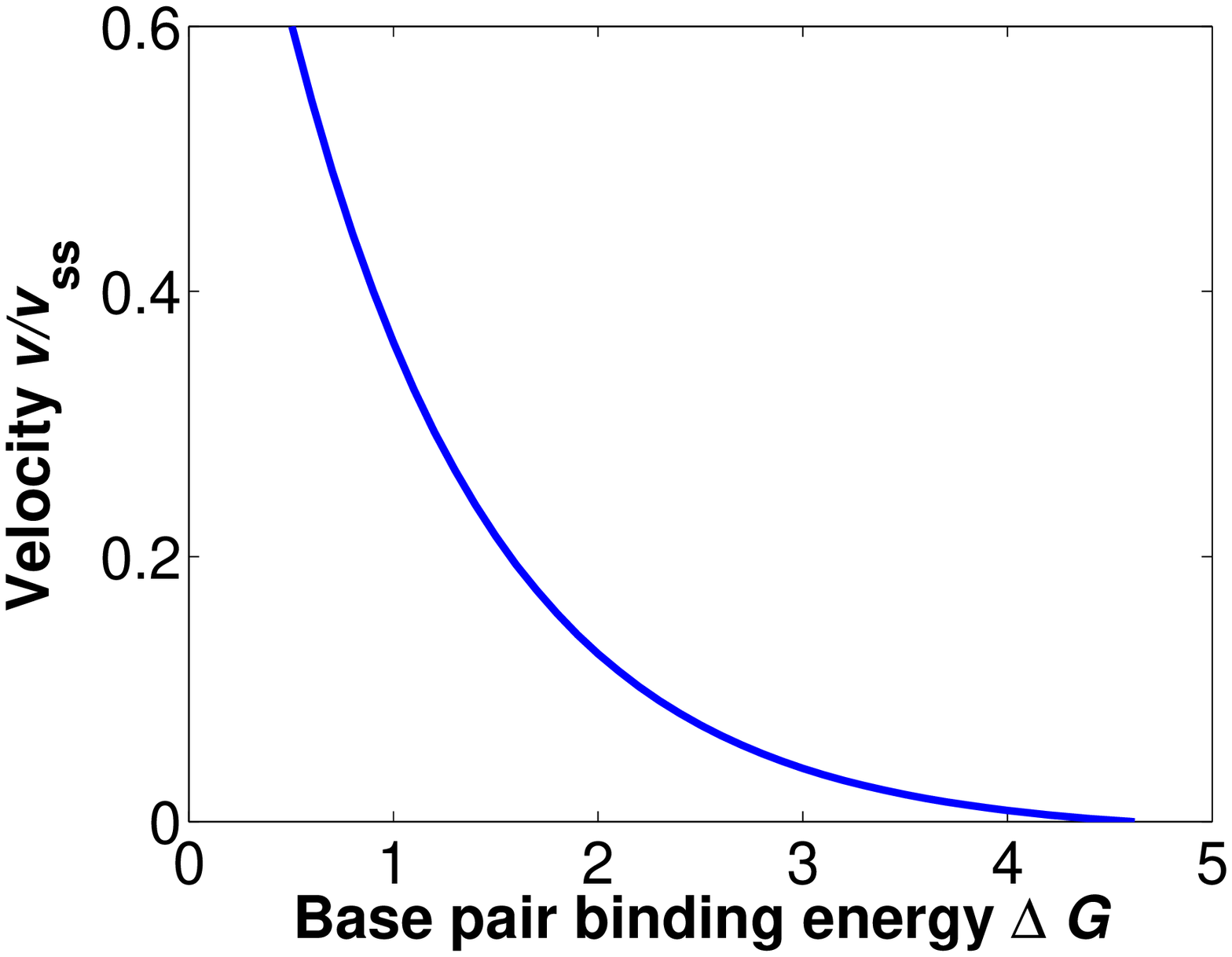}\includegraphics[height=5.0cm]{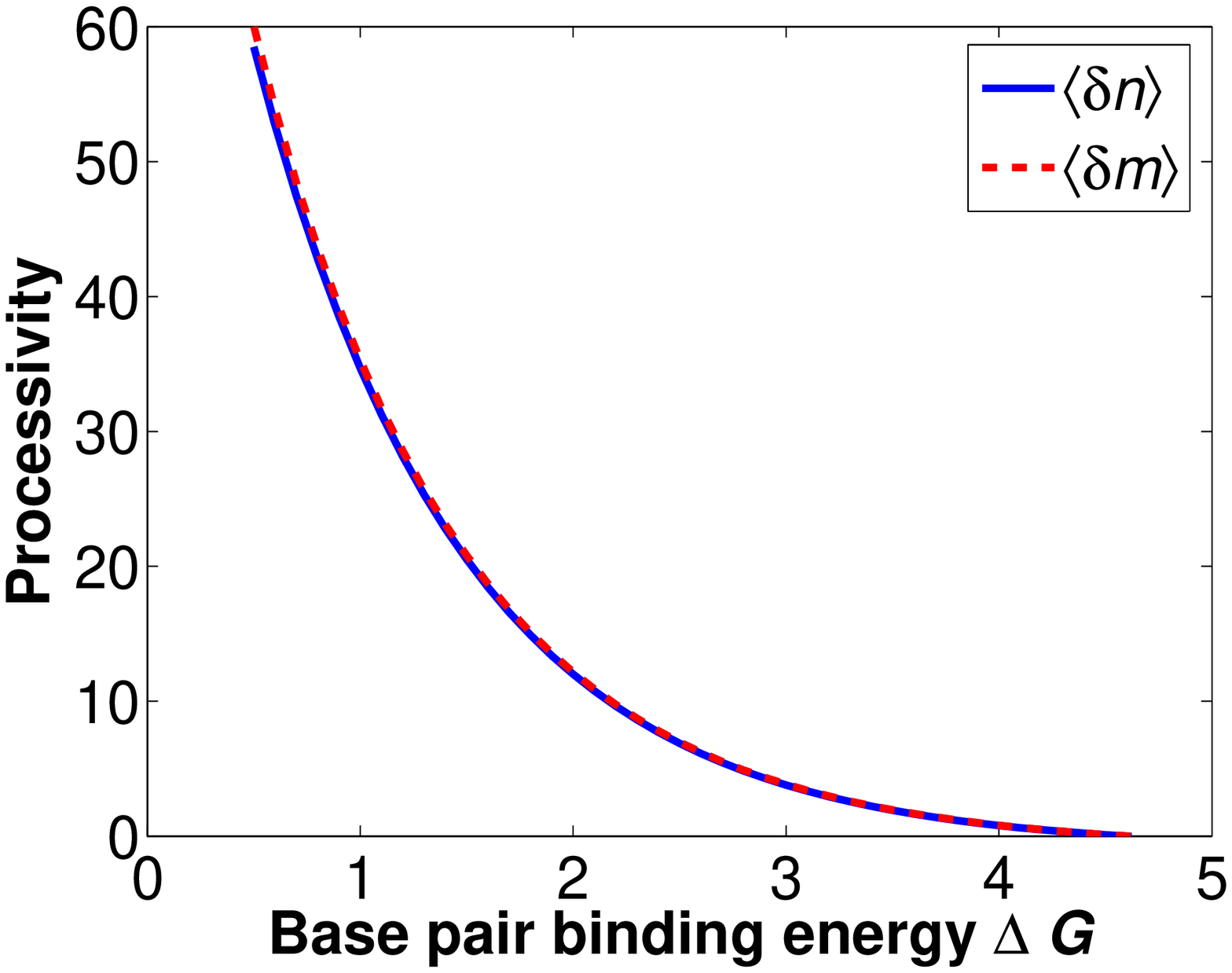}
\caption {Velocity and processivity as a function of the
  base-pairing energy $\Delta G$ for a hard wall-potential.  We varied
  the opening rate of the ss-ds junction, assuming that the closing
  rate is constant.  Therefore $\alpha=\beta e^{-\Delta G}$.  Left,
  unwinding rate $v_{HW}/v_{ss}$ as a function of $\Delta G$. Right,
  translocation and unwinding processivity as a function of $\Delta
  G$.  Parameters are $\beta=7 \times 10^5$ s$^{-1}$, $k^+=1$ base
  s$^{-1}$, $k^-=0.01$ base s$^{-1}$, $\gamma=0.01$ s$^{-1}$, and
  $j_o=1$.  }
\label{gchange} 
\end{figure}

The helicase adjacent to the junction can only move when the junction
opens.  Therefore, the velocity and the processivities depend strongly
on the base pair binding free energy $\Delta G$.  The velocity and the
measures of processivity are displayed in figure \ref{gchange} as a
function of $\Delta G$.  In this calculation, we assume that $\beta$
is constant and $\alpha = \beta e^{-\Delta G}$.  The value $\Delta
G=2$ corresponds to the sequenced-average value for typical genomic
DNA.  For small $\Delta G$, the processivity increases dramatically,
because the binding energy driving NA closing is decreased.

%\begin{figure}[t] 
%\centering
%\includegraphics[height=5.0cm]{varyg2}\includegraphics[height=5.0cm]{varyg1}
%\caption {Velocity and processivity as a function of the
%  base-pairing energy $\Delta G$ for a hard wall-potential.  We varied
%  the opening rate of the ss-ds junction, assuming that the closing
%  rate is constant.  Therefore $\alpha=\beta e^{-\Delta G}$.
% (a) unwinding rate %
% $v_{HW}$ as a function of $\Delta G$, normalized to the value at
% $\Delta G=2$. Right, translocation and unwinding processivity as a
% function of $\Delta G$, normalized to the value at $\Delta G=2$.  The
% strong dependence of the velocity and processivity to $\Delta G$ is a
% signature of passive unwinding. Parameters are $\beta=7 \times 10^5$
% s$^{-1}$, $k^+=1$ bases s$^{-1}$, $k^-=0.01$ bases s$^{-1}$,
% $\gamma=0.01$ s$^{-1}$, and $j_o=1$.  }
%\label{gchange} 
%\end{figure} 

\section{Velocity and processivity of an active helicase}
\label{active}

\subsection{Velocity} 

Active opening can be represented by an interaction potential with
nonzero range between junction and helicase.  For simplicity, we
choose linear potentials characterized by a range of $N$ steps and a
step height $U_o$ (figure \ref{enplot}). The repulsive interaction
between helicase and junction implies that for small separation $j$
within the range of the potential the rates of junction opening and
helicase backward hopping are increased.  The unwinding velocity for
such a potential, relative to the hard-wall case, is
\cite{bet03,bet05}
\begin{equation}
\frac{v_N}{v_{HW}}= \frac{c^{N}(e^{-U_o}-c) + (1-c) e^{-f U_o} 
      (e^{-N U_o}-c^N)}{c^{N}(e^{-U_o}-c) + (1-c) e^{- U_o} 
      (e^{-N U_o}-c^N)}.   
\label{nstep}
\end{equation}  
Note that the helicase cannot increase the unwinding rate beyond a
certain limit. Assuming $0<f<1$, the unwinding rate has an upper bound
$v_N \le c^{-1}v_{HW}$.

\begin{figure}[t] 
  \centering
\includegraphics[height=4.25cm]{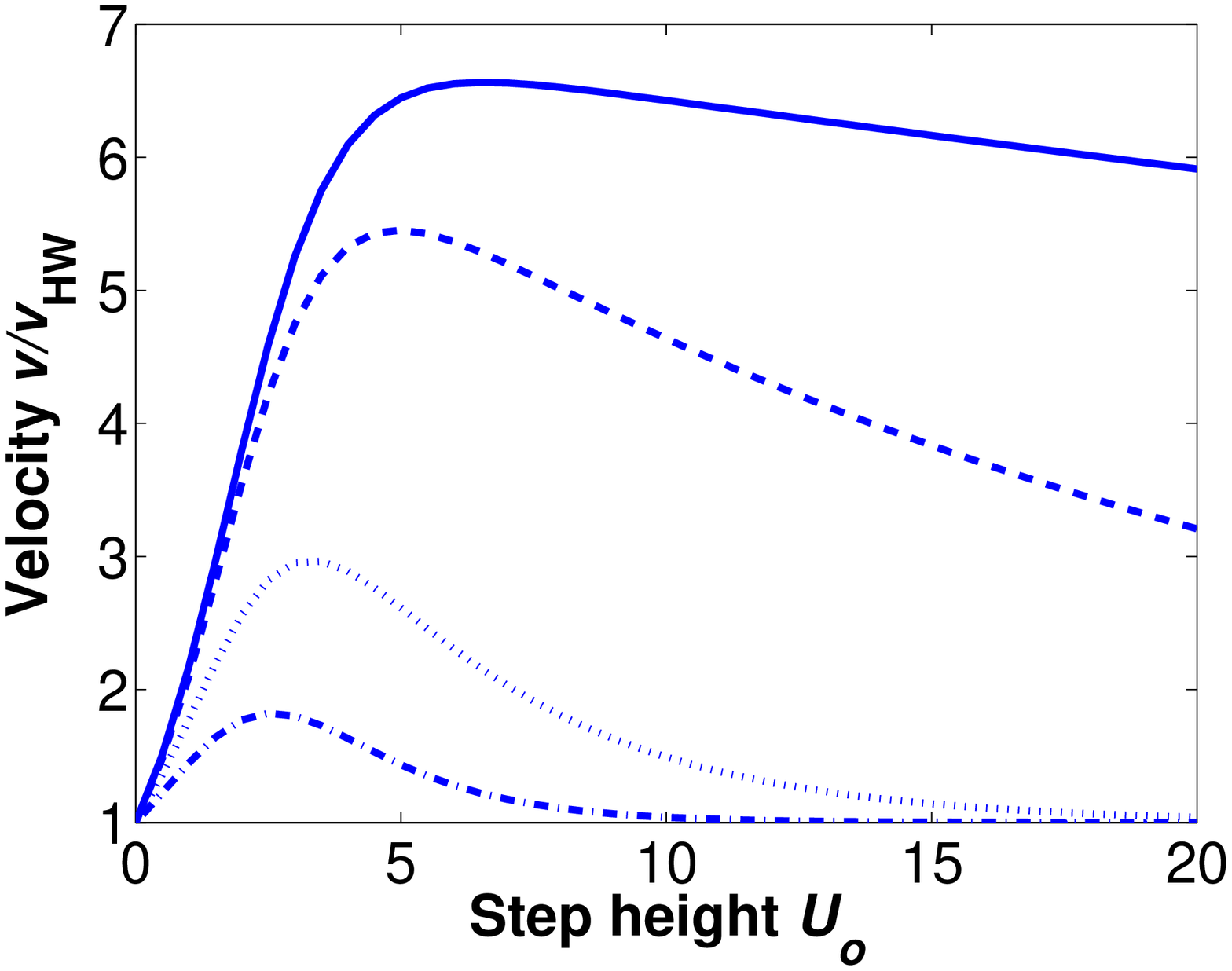}\includegraphics[height=4.25cm]{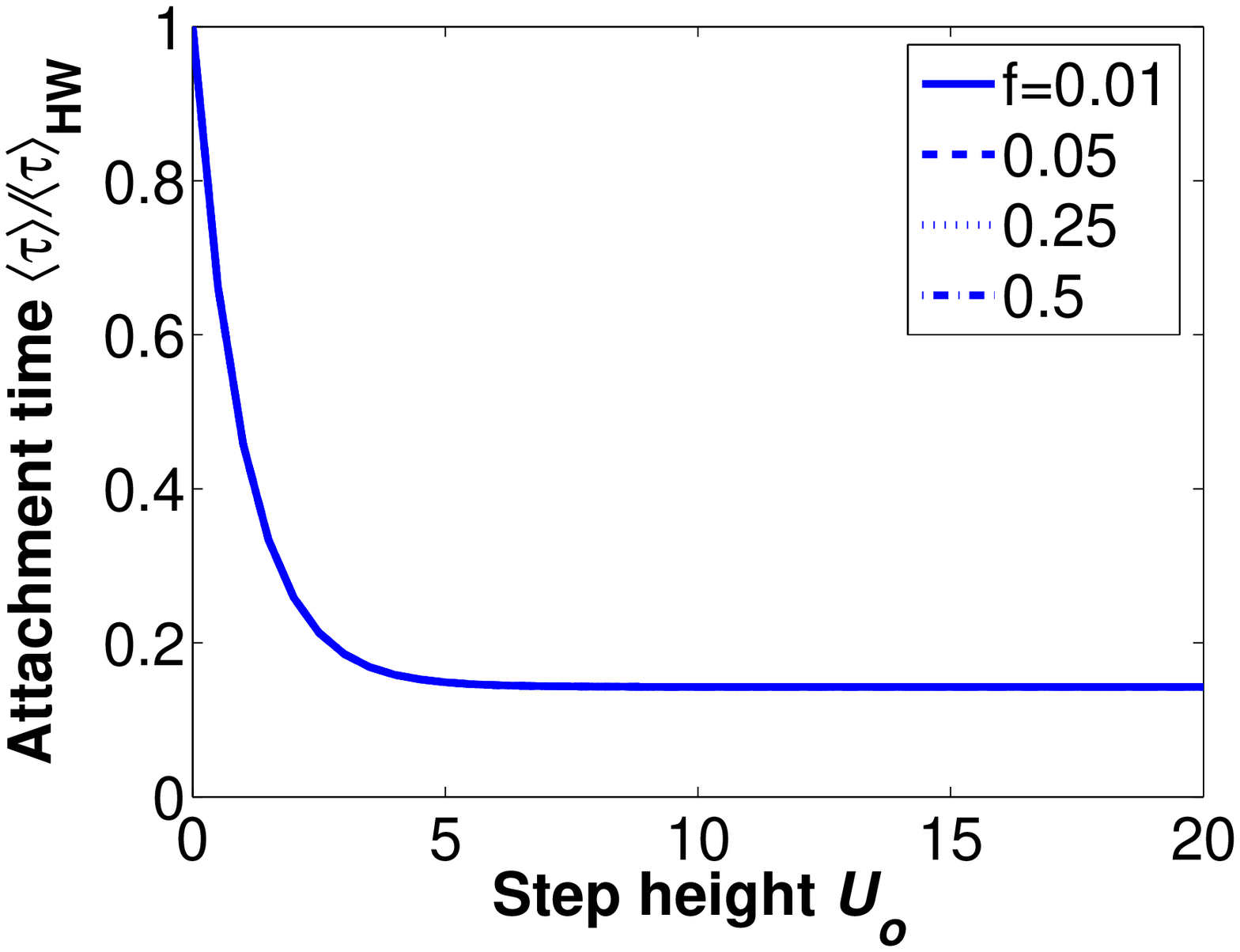}\includegraphics[height=4.25cm]{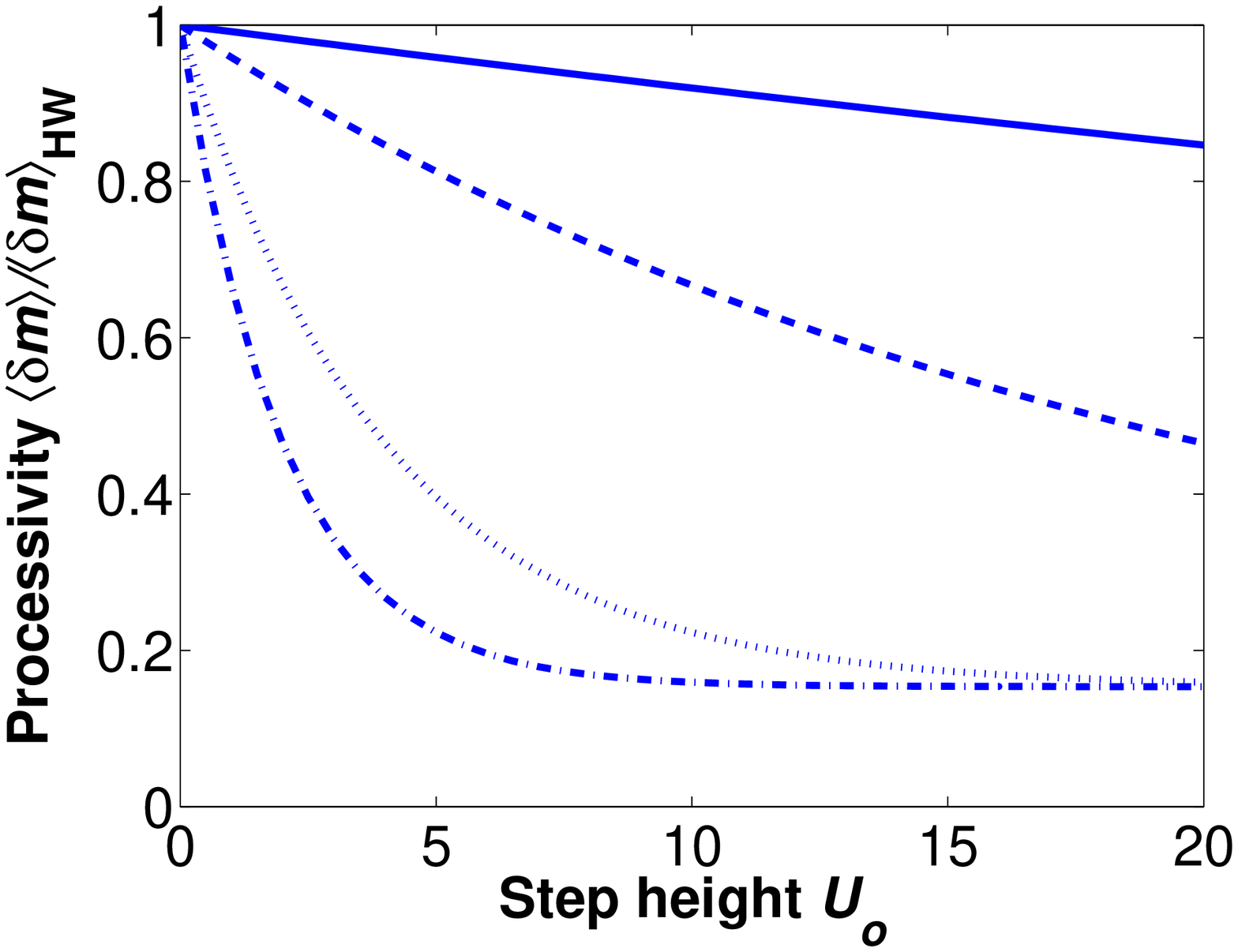}
\caption{Variation of unwinding velocity, average attachment time, and
  unwinding processivity as a function of step height for a one-step
  potential.  The parameters are $\alpha=10^5$ s$^{-1}$, $\beta=7
  \times 10^5$ s$^{-1}$, $k^+=1$ base s$^{-1}$, $k^-=0.01$ base
  s$^{-1}$, and, for the processivity calculation, $\gamma=0.01$
  s$^{-1}$, $j_o=1$, and grid size $M=100$.}
\label{onestepvel} 
\end{figure}

The unwinding velocity for one step ($N=1$) is displayed in figure
\ref{onestepvel}a as a function of the step height $U_o$ for different
values of the parameter $f$.  For small $U_0$, the unwinding rate
increases with increasing step height $U_o$ because the presence of
the step facilitates NA opening.  For large step heights, the
unwinding rate decreases, because the repulsive potential reduces the
rate of helicase forward motion.  The unwinding velocity for different
values of $N$ is shown in figure \ref{varystep}a as a function of
$U_o$.  For increasing $N$, the opening rate becomes more sensitive to
$U_o$ and the maximum occurs at higher values.  For large $N$, the
maximum unwinding rate occurs for $ U_{*} \approx - \ln c \approx
\Delta G$. In this case fastest unwinding occurs when the slope of the
potential matches the base-pairing energy of the NA.

\subsection{Processivity for active unwinding}

Active unwinding results in a decrease of the mean attachment time, as
compared to ss translocation or passive opening for which $\langle
\tau \rangle =\gamma^{-1}$.  This is a consequence of the repulsive
interaction potential which leads to increased unbinding rates for
small $j$.  The dependence of the average attachment time on the step
height is displayed in figure \ref{onestepvel}b for $N=1$.  For a
potential with one step, $\langle \tau \rangle $ decreases rapidly
with step height to a plateau value $\approx 0.2 \gamma^{-1}$. This
decrease is independent of $f$, since the unbinding rate does not
depend on the barrier between states of different $j$. Note that the
limits $U_o \to 0$ and $U_o \to \infty$ do not give the same result
for the processivity because the initial conditions are different:
$U_o \to 0$ corresponds to a hard-wall potential with $j_o=1$, while
$U_o \to \infty$ corresponds to a hard-wall potential but with
$j_o=0$.

\begin{figure}[t] 
  \centering
  \includegraphics[height=4.25cm]{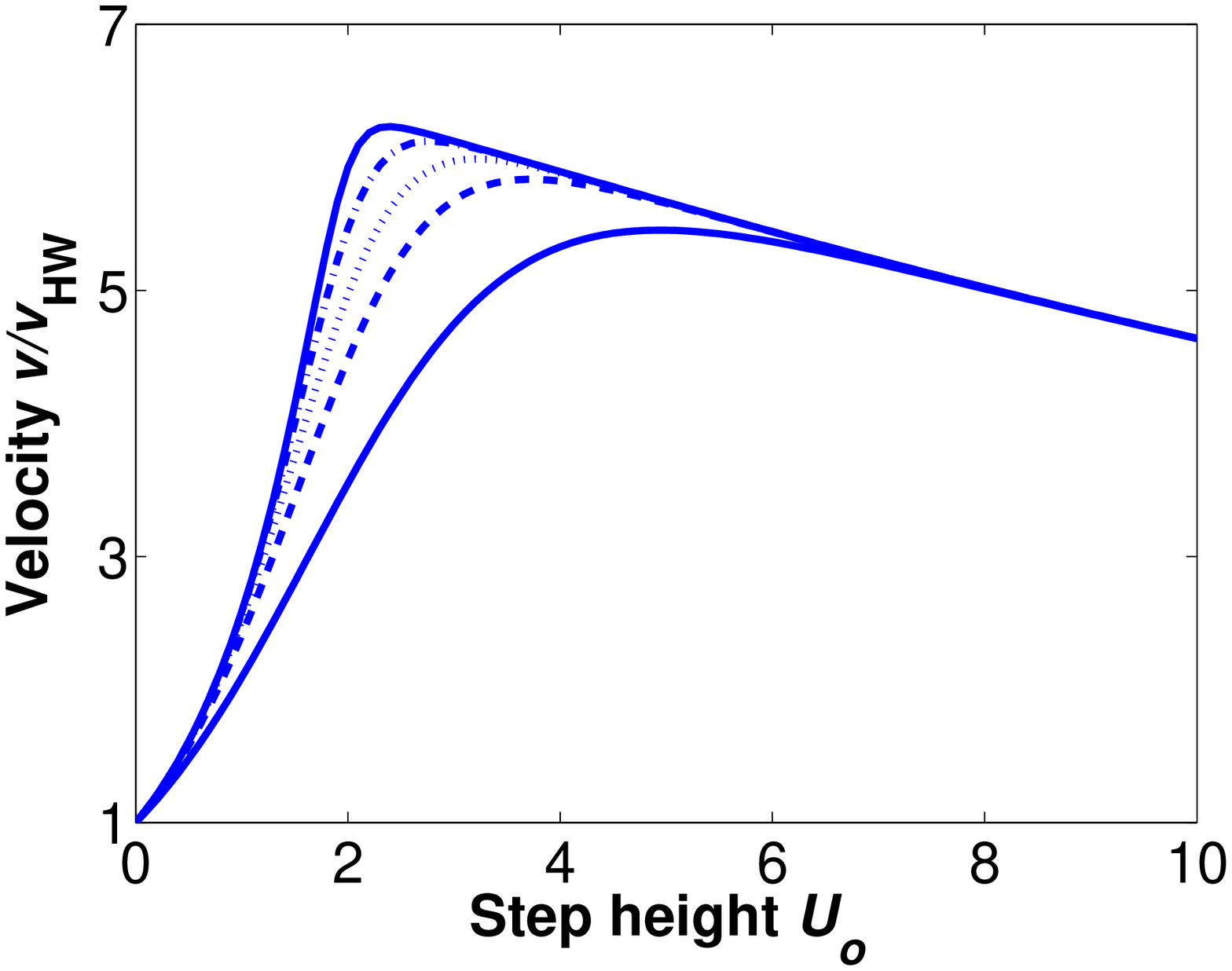}\includegraphics[height=4.25cm]{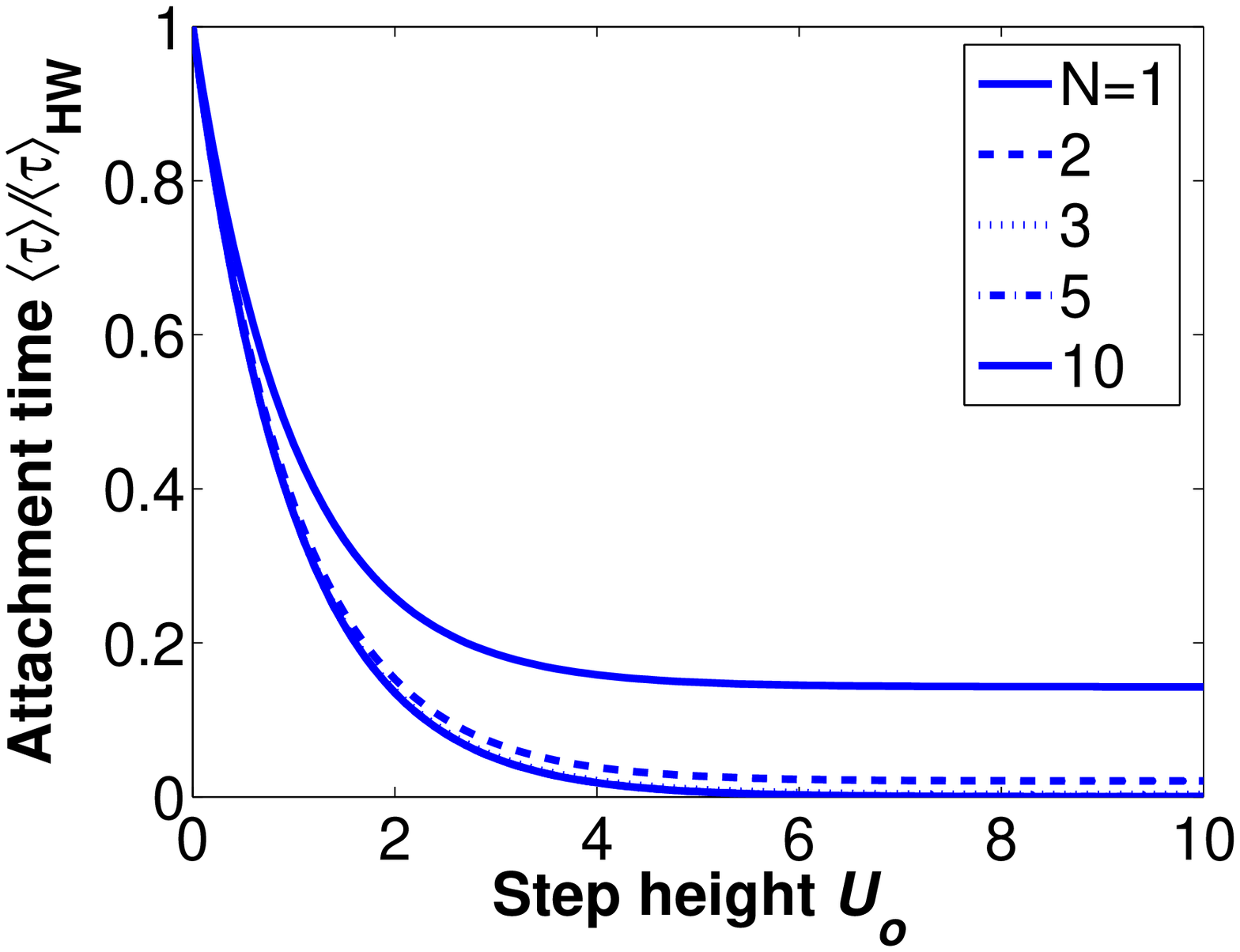}\includegraphics[height=4.25cm]{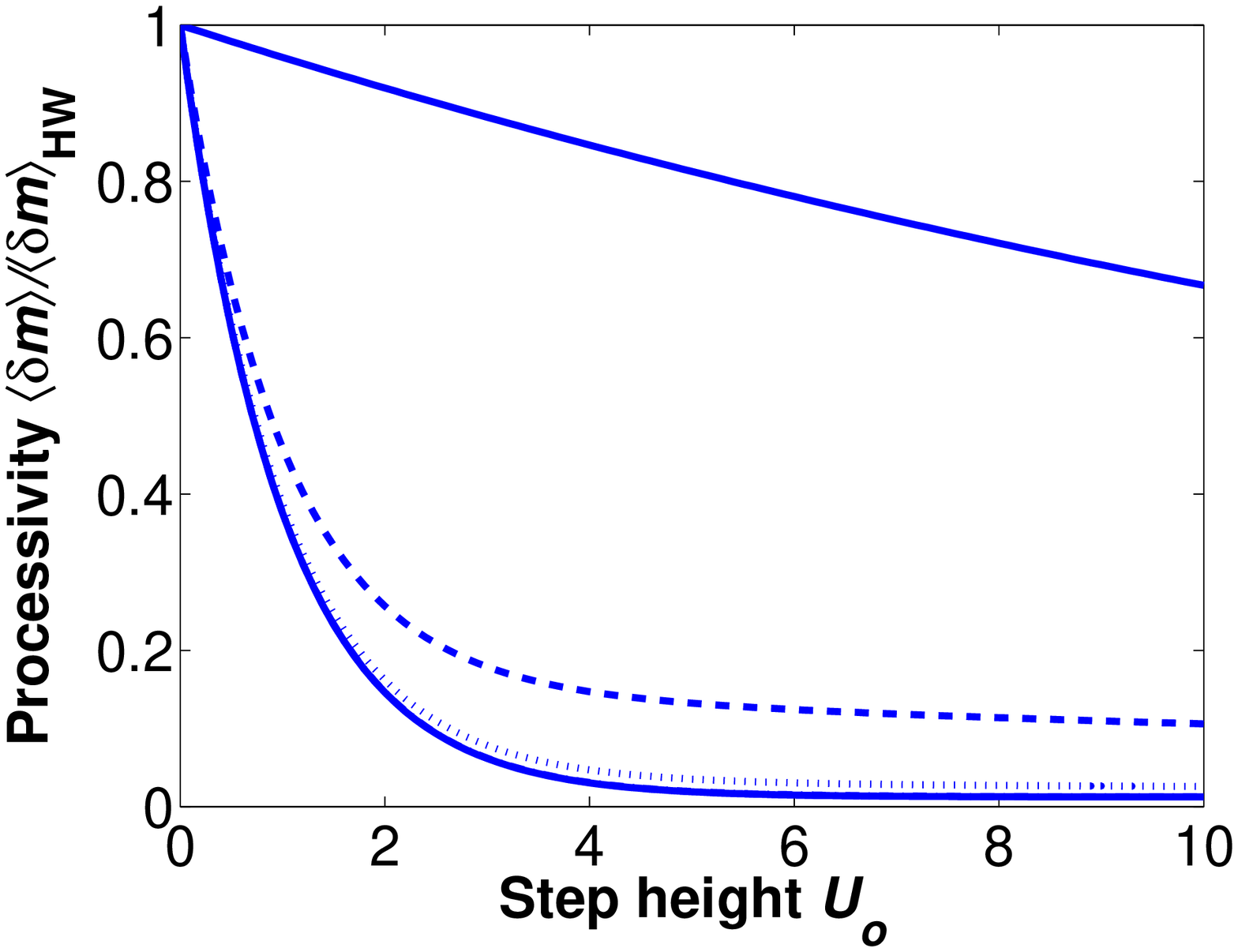}
\caption{Variation of velocity, average attachment time, and unwinding
  processivity with step height, for potentials with varying numbers
  of steps. The parameters are $f=0.05$, $\alpha=10^5$ s$^{-1}$,
  $\beta=7 \times 10^5$ s$^{-1}$, $k^+=1$ base s$^{-1}$, $k^-=0.01$
  base s$^{-1}$, and, for the processivity calculation, $\gamma=0.01$
  s$^{-1}$, $j_o=1$, and $M=100$.}
\label{varystep} 
\end{figure}

The decrease of $\langle \tau \rangle $ for increasing $U_o$ is
similar but even more pronounced for potentials with longer range, as
shown in figure \ref{varystep}b. For $N\ge 2$, the average attachment
time rapidly decreases to $<0.01 \gamma^{-1} $. Decreased attachment
time during unwinding (as compared to ss translocation) is a strong
signature of active unwinding.

Because the attachment time decreases with increasing step height, the
translocation and unwinding processivities also tend to decrease.
However, for a one-step potential $\langle \delta n \rangle $ and
$\langle \delta m \rangle $ decrease slowly for increasing step height
when $f$ is small (figure \ref{onestepvel}c). This occurs because the
rapid increase in the velocity approximately cancels the decrease in
attachment time, leading to a processivity that is relatively
insensitive to step height. For larger numbers of steps, a decrease in
$\langle \delta m \rangle $ occurs as $U_o$ increases (figure
\ref{varystep}c).  These curves are approximately exponentially
decreasing functions, with a decay constant determined by $f$. Small
$f$ means that the processivity varies slowly in the biologically
plausible range between $0-20$ $k_B T$. Note that above $U_o=20$,
little or no increase in the velocity can be achieved by increasing
the step size.

\subsection{Dependence on base-pair binding energy}
\label{activeg}

\begin{figure}[t] 
\centering
\includegraphics[height=4.25cm]{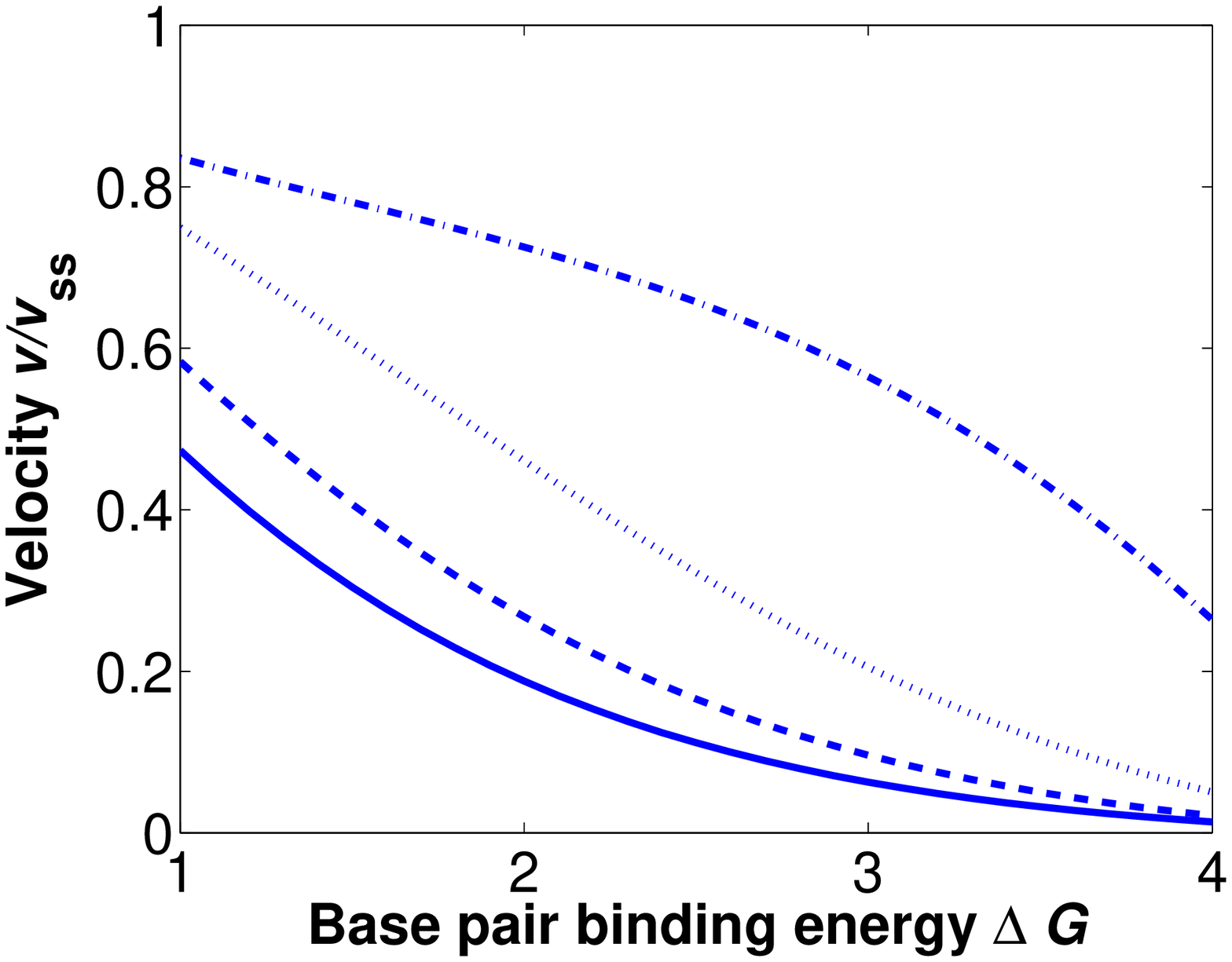}\includegraphics[height=4.25cm]{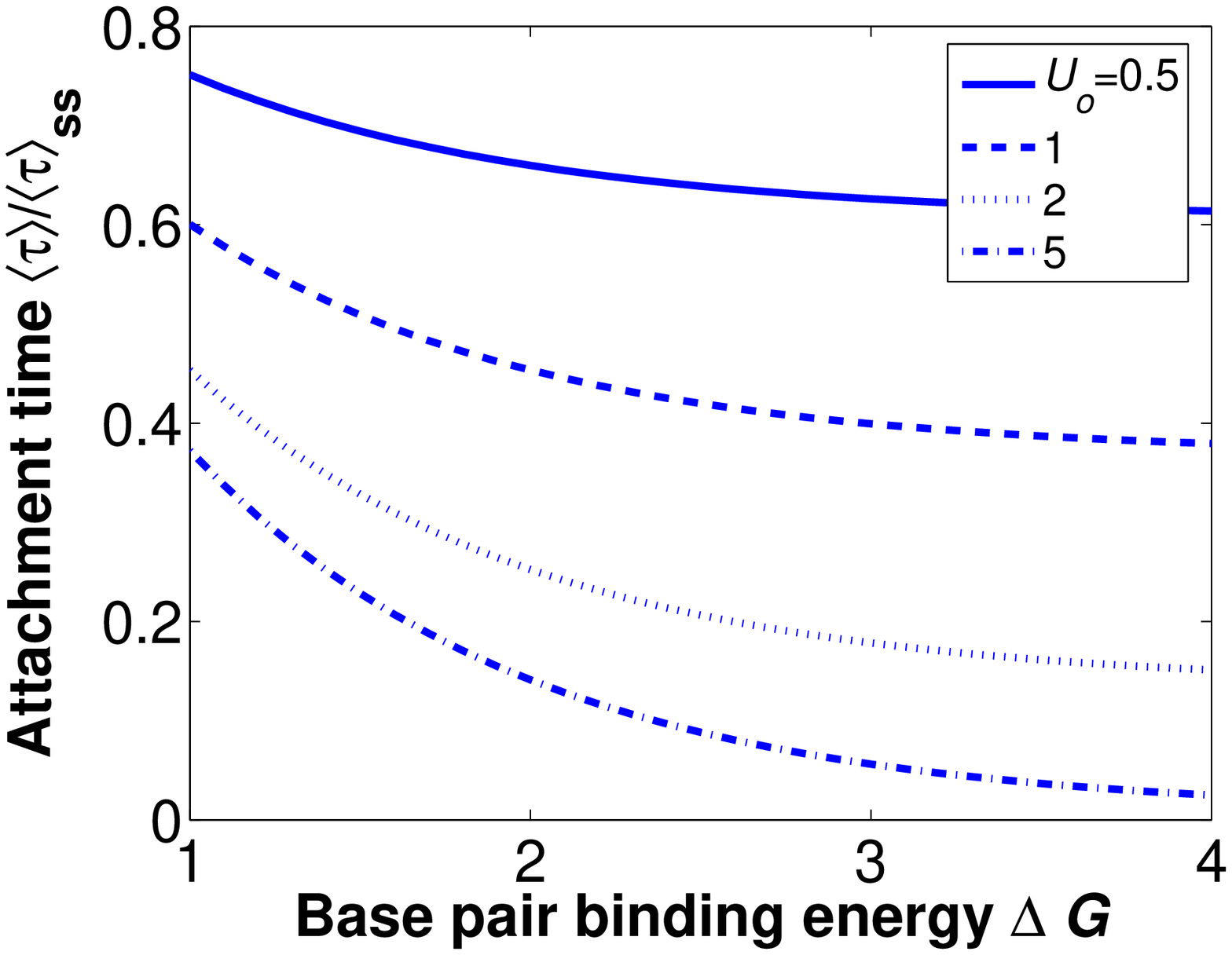}\includegraphics[height=4.25cm]{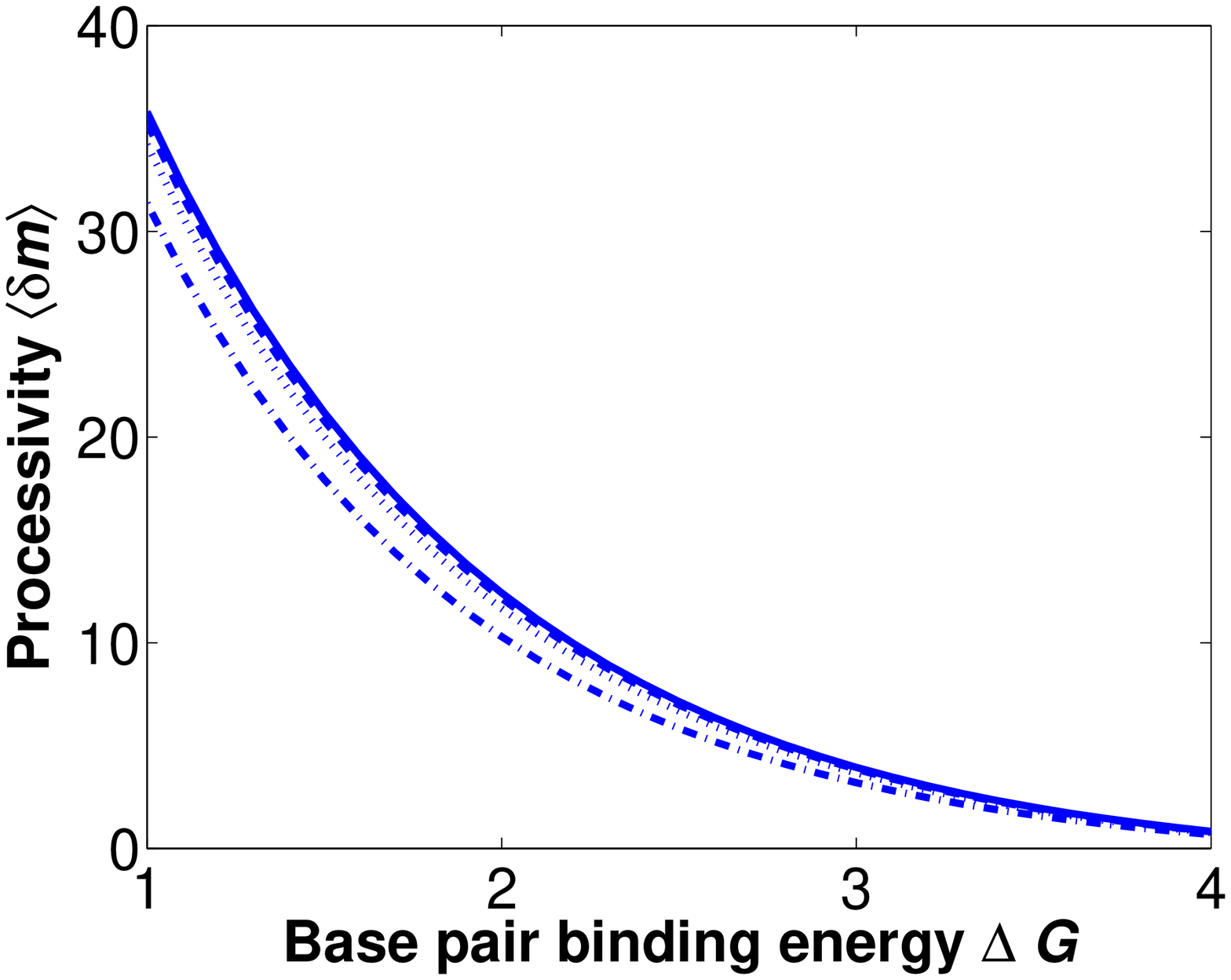}
\caption
{Unwinding velocity, attachment time, and unwinding processivity
  \textit{vs.} free-energy per base pair for a one-step potential. We
  fix the closing and therefore $\alpha=\beta e^{-\Delta G}$.  The
  parameters are $f=0.05$, $\beta=7\times 10^5$ s$^{-1}$, $k^+=1$ base
  s$^{-1}$, $k^-=0.01$ base s$^{-1}$, and, for the processivity
  curves, $\gamma=0.01$ s$^{-1}$, $j_o=1$, and $M=100$.}
\label{force} 
\end{figure}

The velocity and attachment time are sensitive to the mean free energy
per base pair $\Delta G$. We assume that the rate $\beta$ is constant
and $\alpha = \beta e^{-\Delta G}$. For simplicity, we show results
only for a potential with $N=1$. The dependence of $v$ and $\langle
\tau \rangle $ on $\Delta G$ depends on the step height $U_o$.  For a
hard-wall potential, the velocity varies exponentially as a function
of $\Delta G$. As $U_o$ increases, the velocity becomes less sensitive
to $\Delta G$ (figure \ref{force}a). Thus for $U_o=5$ the displayed
curve decreases slowly with increasing $\Delta G$.  The attachment
time $\langle \tau \rangle $ shows the opposite trend: for a hard-wall
interaction potential, the attachment time is independent of $\Delta
G$. The attachment time is also weakly dependent on $\Delta G$ for
small $U_o$, while for larger $U_o$, $\langle \tau \rangle$ varies
more rapidly (figure \ref{force}b). The processivity $\langle \delta m
\rangle $ as a function of $\Delta G$ exhibits the same behavior
independent of $U_o$ (figure \ref{force}c), because the variation in
velocity and attachment time as a function of step height
approximately cancel. Therefore, measurements of velocity and
attachment time have greater power to elucidate aspects of the
potential.

%\begin{figure}[t] 
%\centering
%\includegraphics[height=4.25cm]{varygv}\includegraphics[height=4.25cm]{varygt}\includegraphics[height=4.25cm]{varygm}
%\caption
%{Unwinding velocity, attachment time, and unwinding processivity
%  \textit{vs.} free-energy per base pair for a one-step potential. We
%  fix the closing rate $\beta=7\times 10^5$ and therefore
%  $\alpha=\beta e^{-\Delta G}$. Left,unwinding velocity. Middle,
%  attachment time. Right, unwinding processivity. All curves show the
%  values normalized to the value at $\Delta G=2$.  The parameters are
%  $f=0.05$, $\beta=7\times 10^5$, $k^+=1$, $k^-=0.01$, and, for the
%  processivity curves, $\gamma=0.01$, $j_o=1$, and $M=100$.}
%\label{force} 
%\end{figure} 

\section{Discussion}
\label{conclusion}

In this paper, we have extended a model for the unwinding of
double-stranded nucleic acids by helicases to allow calculation of
helicase processivity. An interaction potential describes how the
helicase and NA ss-ds junction affect each other.  We assume that the
local unbinding rate of the helicase depends exponentially on the
interaction potential.  The mean attachment time, the translocation
processivity, and the unwinding processivity depend on initial
conditions: if the position where the helicase binds to the ssNA is
varied, the processivity changes.  Therefore, in different
experimental situations different values of the processivity could be
obtained.

A hard-wall interaction potential describes passive unwinding.  In
this case, the unwinding velocity $v_{HW}$ is significantly slower
than the single-strand translocation rate of the motor far from the
ss-ds junction.  The average attachment time for the hard-wall
potential is $\langle \tau \rangle = \gamma^{-1}$ since the unbinding
rate $\gamma$ is independent of position.  In other words, the mean
attachment time of a passive helicase is the same whether the helicase
translocates on ssNA or unwinds dsNA.  For initial conditions where
the helicase and junction are adjacent to each other, $\langle \delta
m \rangle \approx v_{HW} \langle \tau \rangle$---the translocation
processivity is approximately the unwinding velocity times the
attachment time.  The translocation processivity is lower for a
passive helicase during unwinding (relative to ss translocation); this
decrease occurs solely because of the decrease in velocity of the
helicase, while the average attachment time is unchanged.

We represent active unwinding by a linearly increasing potential of
finite range, characterized by the step height $U_0$ and the number of
steps $N$.  A range of up to 10 bases is motivated by the typical size
of a helicase monomer.  Indeed, structures determined by X-ray
diffraction suggest that a helicase can interact with 5-10 bases in
the ds region of the ss-ds junction
\cite{sub96,korol97,kim98,korol98,vel99,machius99,sin00}.  We find
that for an active helicase the unwinding velocity can approach the
single-strand translocation rate of the motor. In other words, the
helicase can unwind as fast as it translocates on ssNA.

In contrast to the passive case, the attachment time of an active
helicase is shorter during unwinding than during ss translocation.
Decreased attachment time during unwinding is a general property of
actively opening helicases in our model. Our representation of the
unbinding rate assumes that the helicase is bound in a single
potential well with a barrier separating the bound state from the
unbound state.  This model is an approximation, because the helicase
exists in different biochemical states throughout the ATP hydrolysis
cycle. Each state might experience a different free-energy barrier to
unbinding. In principle, if the relative time spent in each of these
states changes (for example, as the ATP concentration is varied), the
unbinding rate might vary as well. However, experimental evidence
suggests that our simplified model is a good approximation.
Single-molecule measurements on RecBCD\cite{bianc01} and
UvrD\cite{dessin04} find that the average attachment times of these
helicases are independent of ATP concentration. Therefore, the
assumption that the potential well for a bound helicase is independent
of hydrolysis state is consistent with experiments.

In our model, an actively unwinding helicase shows a decrease in
attachment time. Previous work has suggested that high helicase
processivity may require two helicase-NA binding sites---one on the
ssNA, and another on the dsNA\cite{loh96,brigg05,rodrig05}. In this
mechanism, a helicase may unbind from the ssNA but remain bound at the
dsNA site.  The opportunity for dsNA binding is absent when the
helicase translocates on ssNA. This effect can be incorporated in our
description by introducing an extra binding energy and consequently
reduced detachment rate if the helicase is close to the junction.

Both the velocity of unwinding and measures of processivity are
sensitive to $\Delta G$, the average free energy of NA base-pair
opening.  The value of $\Delta G$ can be controlled in single-molecule
experiments where tension is applied to the ends of the NA,  by the
concentration of ss-binding proteins in the buffer solution, and by
varying the base composition of the NA.

Measurements of helicase velocity and processivity as a function of
$\Delta G$ can provide information about the interaction between
helicase and NA ss-ds junction.  Because the rate of passive unwinding
is determined by the opening probability of the base pair at the
junction, the velocity and the processivity decrease rapidly as
$\Delta G$ increases. For active opening, the unwinding velocity as a
function of $\Delta G$ depends sensitively on the step height: while
for larger values of $U_o$ the velocity depends only weakly on $\Delta
G$, this dependence becomes strong for small $U_0$.  The average
attachment time $\langle \tau \rangle $ behaves differently.  For a
hard-wall interaction potential, the attachment time is independent of
$\Delta G$. For active unwinding, the attachment time depends strongly
on $\Delta G$ for larger $U_o$ and tends to decrease for increasing
$\Delta G$.  The effects of $\Delta G$ on velocity and attachment time
approximately cancel when computing the processivities, so that the
behavior of $\langle \delta m \rangle $ as a function of $\Delta G$ is
similar for different step heights. 
Galletto \textit{et al.}\ found that the unwinding rate of DnaB
helicase depends on the DNA GC content\cite{gallet04}. Varying the
fraction of GC versus AT base pairs changes the average $\Delta G$ of
the NA template. The dependence on GC content found in these
experiments would be expected in our model for an interaction
potential with a small (or zero) step height.

For the helicase PcrA, crystal structures suggest that the protein
binds both the ss and dsDNA and distorts the double helix
\cite{vel99}.  If the protein residues which have been proposed to
interact with dsDNA are mutated, the mutant proteins hydrolyze ATP in
the presence of ssDNA at a rate similar to the wild-type protein, but
unwind dsDNA 10-30 times more slowly than wild type\cite{sou00}. In
the language of our analysis, the mutations may alter the interaction
potential between junction and helicase such that it resembles the
passive (hard-wall) case.  In our description, starting from a
situation of active opening with optimally chosen step height and step
number to a passive case with hard-wall potential typically leads to a
decrease of the unwinding rate by a factor of $c^{-1}\approx7$.
Further altering the potential to an attractive linear
potential\cite{bet03} with a well depth of 2 $k_BT$ decreases the
unwinding rate by a factor of 35.  Small changes to the interaction
potential can thus cause the unwinding rate to vary by a large factor.

In single-molecule experiments on UvrD helicase, Dessinges \textit{et
  al.}\ observed the unwinding of DNA molecules by UvrD at different
forces\cite{dessin04}. In these experiments, the end-to-end extension
of a tethered DNA molecule changes with time as a result of the
transformation of dsDNA into ssDNA by the helicase.  These
single-molecule experiments found that unwinding can be induced by
UvrD monomers.  However, bulk experiments have suggested that helicase
activity requires UvrD dimers\cite{maluf03,maluf03b,fisch04b}, even
though UvrD monomers have a large processivity (2400 bases) when
translocating on ssDNA\cite{fisch04b}.  The reasons for this
difference are not currently understood.  The velocity of unwinding
was shown by Dessinges \textit{et al.} to depend on ATP
concentration\cite{dessin04}.  In addition, events were observed where
the DNA slowly closed at an ATP-dependent rate.  It was suggested that
these events occur when a UvrD molecule bound near the ss-ds junction
switches from one DNA single strand to the other.  Because the two
single strands have opposite polarity, the helicase might move away
from the junction after strand switching.  In this interpretation, the
re-zipping events provide information on ss translocation of UvrD. The
re-zipping velocity would be expected to be greater than or equal to
the ss translocation rate of UvrD, because the energetically favorable
re-annealing of the two DNA strands behind the helicase might
accelerate the protein's motion.
%unbinding rates during ss trans by Lohman: 0.08 s^{-1}
%unbinding rates during rezipping: 0.034 s^{-1}; both at 500 micro
%molar ATP concentration.
%unbinding rates during ds unwind by Lohman:
%note: unbinding occurs by dimers; monomers can't unwind even 18 bp
%lengths in their experiments
%unbinding rates during ds unwind by Bensimon: 0.36 s^{-1}
%for translocation rate of 41.3 s^{-1}, 3 times larger than in bulk

We can relate three key measurements of UvrD motion by Dessinges
\textit{et al.}\cite{dessin04} to the behavior of active unwinding
discussed here.  First, the experiments show that the re-zipping
velocity and the unwinding velocity are comparable in certain
situations: the re-zipping velocity is approximately 15\% larger than
the unwinding rate at 35 pN applied force. Second, the effective
detachment rate is about 10 times larger during unwinding than during
re-zipping (at 35 pN).    Third, the measured unwinding
velocity is only weakly dependent on $\Delta G$.  The velocity was
measured for 3 pN and 35 pN applied force on the DNA.  The first case
corresponds to an increase of $\Delta G$, while in the second case
$\Delta G$ decreases\cite{dessin04}.  If we interpret re-zipping
events as approximating the ss translocation behavior of UvrD, all
three of these experimental observations are consistent with our
description of an active helicase. 

The comparison of these data to bulk measurements reveals agreement in
the unbinding rate during ss translocation but not in the unwinding
rate. The unbinding rate of UvrD monomers translocating on ssDNA
measured in bulk by Fischer \textit{et al.} is comparable to (2 times
larger than) the unbinding rate measured by Dessinges \textit{et al.}
for UvrD monomers during re-zipping events\cite{fisch04b}. Fischer
\textit{et al.} found that the UvrD unwinding rate was approximately 3
times slower than the ss translocation rate\cite{fisch04b}. However,
the bulk experiments of Fischer \textit{et al.} observed unwinding
only by UvrD dimers, whereas the single-molecule experiments of
Dessinges \textit{et al.} observed unwinding by UvrD monomers.
Therefore the unwinding data may not be directly comparable.

%Dessinges \textit{et al.}\ present evidence that the physical step
%size of UvrD is 6 bases\cite{dessin04}, which is comparable to the
%step size of 4-5 bases estimated by fitting bulk measurements to a
%kinetic model of UvrD motion\cite{fisch04b}.  In our model, we have
%assumed that the step size of the helicase is 1 base.  However, we can
%estimate the effects of a larger step size if we renormalize the size
%of junction-opening fluctuations to equal the helicase step size. In
%other words, a passive helicase which takes 6-base forward steps would
%have to wait for a thermal fluctuation to open 6 bp at the junction
%before it could advance. In this case, the maximal hard-wall velocity
%becomes
%\begin{eqnarray}
%  v_{HW} &= 6 k^+ \left( \frac{\alpha}{\beta} \right)^6, \nonumber \\
%&\approx 5 \times 10^{-5} k^+.
%\end{eqnarray}
%If a helicase takes multiple-base steps, passive unwinding is many
%orders of magnitude slower than ss translocation. Therefore it is
%unlikely that helicases which take multiple-base steps unwind
%passively. Data which show that a helicase unwinds more slowly than it
%translocates on single strands\cite{jeong04} do not necessarily
%indicate a passive mechanism, if the helicase takes multiple-base
%steps.

Our physical theory describes how helicase unwinding velocity and
processivity depend on the interaction potential between the helicase
and the ss-ds junction.  This type of theory includes more detail then
models based on kinetic states and transitions between them, while
neglecting many details present in all-atom simulation models.
Physical descriptions of the type we describe here complement other
modeling frameworks. Simplifying some aspects of the system---for
example, our choice to neglect the biochemical states which
occur during helicase ss translocation---allows us to focus on the
coupling between helicase translocation and NA unwinding.  This level
of detail may allow one to relate observed helicase behavior to simple
physical mechanisms, and thereby gain more insight into the parameters
which are important for helicase behavior.  For example, experiments
on UvrD suggest that this helicase unbinds more rapidly when unwinding
dsDNA that during ss translocation\cite{dessin04}. Our model gives a
physical picture for why this might happen: an interaction potential
corresponding to active opening naturally leads to accelerated
helicase unbinding.

Our simple description neglects several effects.  We ignore
deformations of the NA strand, such as bending and torsion, and treat
the strand as a rigid structure.  The helicase is described by forward
and backward rates only; we neglect the details of the protein's
biochemical states.  In addition, we ignore the effects of the NA base
sequence on opening. These effects are believed to be weak for most
helicases\cite{loh96}, although sequence dependence has been
demonstrated for Rho\cite{walmac04} and DnaB\cite{gallet04} helicases.
Recent work by Kafri, Lubensky and Nelson\cite{kaf04} shows that a
motor protein which translocates on a random track can show
interesting behavior near the stall force.

\ack We thank D. Lubensky, T.  Perkins, and J. Prost for helpful
discussions. MDB acknowledges funding from the Council on Research and
Creative Work of the University of Colorado.

\bibliography{helicase}
\bibliographystyle{unsrt}

\end{document}